\documentstyle[amstex,amssymb,epsfig,12pt]{article}

\input{tcilatex}

\begin{document}

\bigskip \bigskip 
\begin{titlepage}
\bigskip \begin{flushright}
hep-th/0207123\\
WATPPHYS-TH02/04
\end{flushright}


\vspace{1cm}

\begin{center}
{\Large \bf {Nutty Bubbles}}\\
\end{center}
\vspace{2cm}
\begin{center}
 A.M. Ghezelbash$^{\dagger}${%
\footnote{%
EMail: amasoud@@sciborg.uwaterloo.ca}} and R. B. Mann$^\ddagger$
\footnote{
EMail: mann@@avatar.uwaterloo.ca}\\
$^{\dagger,\ddagger}$Department of Physics, University of Waterloo, \\
Waterloo, Ontario N2L 3G1, Canada\\
$^{\dagger}$Department of Physics, Alzahra University, \\
Tehran 19834, Iran\\
\vspace{1cm}
\today\\
\end{center}

\begin{abstract}
We investigate the various time-dependent bubble spacetimes that can be 
obtained from double analytic continuation of asymptotically locally flat/AdS spacetimes
with NUT charge. We find different time-dependent explicit solutions of 
general relativity from double analytic continuations of Taub-Nut(-AdS) 
and Kerr-Nut(-AdS) spacetimes. One solution in particular has Milne-like evolution
throughout, and another is a NUT-charged generalization of the AdS soliton. 
These solutions are all four dimensional. In certain situations the NUT charge induces
an ergoregion into the bubble spacetime and in other situations it quantitatively 
modifies the evolution of the bubble, as when rotation is present. 
In dimensions greater than four,  no consistent bubble solutions are found that 
have only one timelike direction.
\end{abstract}
\end{titlepage}\onecolumn

\begin{center}
\end{center}

\section{Introduction}

One of the key open questions in the pursuit of a quantum theory of gravity
is that of understanding its behaviour in time-dependent scenarios. \ In the
context of string theory this has recently led to intensive investigation of
asymptotically de Sitter spacetimes to see what the prospects are for
developing a de Sitter/CFT correspondence \cite{dsCFT}. \ Another line of
approach has been to construct simple time-dependent solutions that provide
(at least to leading order) consistent backgrounds for string theory. This
has been carried out in asymptotically flat spacetimes, and recently
extended to include the asymptotically anti de Sitter (AdS) case. \ This
latter situation is of interest since the AdS/CFT correspondence conjecture
could be employed to relate the time-dependence to the behaviour of the
non-perturbative field theory dual. In so doing one might be able to avoid
(or at least ameliorate) difficulties that arise in understanding the
behaviour of string theory on time-dependent backgrounds.

In this paper we extend the search for time-dependent backgrounds to
asymptotically locally flat/AdS spacetimes. \ Such spacetimes have a form of
magnetic mass referred to as a NUT charge. Their Euclideanized solutions
have been a subject of study in recent years \cite
{NUTpapers,MannNUT,Mannnutrot,mynut}, in part because in the asymptotically
locally AdS case they are relevant to 2+1 dimensional \ ``exotic'' conformal
field theories that live on the world volume of M2-branes (and closely
related theories ), after placing them on squashed three spheres. \ We
consider here the multiple analytic continuation of such spacetimes to form
time-dependent bubble spacetimes with nut charge. \ Such ``nutty bubbles'' \
form interesting time-dependent backgrounds that are asymptotically locally
flat/AdS, extending the variety of time-dependent cases previously studied 
\cite{Aha,Bir,Bala,cai}. The stability of these time-dependent bubble
solutions has also recently been investigated \cite{gib}.

We consider in this paper the different (even-dimensional) Taub-Nut-AdS
spacetimes and study their different analytic continuations. After a brief
overview of bubble solutions obtained from spherically symmetric spacetimes
in section 2, we go on in section 3 to find three bubble solutions in the
four dimensional case, all time-dependent. Two of them are obtained from
analytic continuation of Taub-Nut-AdS spacetime and the last one is inspired
from the first two bubble solutions. As all of these `nutty' bubbles are\
four dimensional, no issue of classical gravitational instability arises in
contrast to the situation considered in\ ref.\cite{gib}. We then compute the
boundary stress-energies associated with such solutions in section 4. \ In
section 5 we consider double analytic continuations of four dimensional
topological NUT-charged spacetimes. We find here some new static bubble-type
spacetimes, and an interesting new time-dependent solution. We find that the
presence of a NUT charge induces an ergoregion, in which timelike curves
must have non-zero momenta in a compact direction. In section 6 we consider
double analytic continuations of four dimensional NUT-charged spacetimes
with a flat subspace and find a new static bubble-type spacetime that is a
NUT-charged generalization of the AdS soliton, as well as an interesting new
time-dependent solution: it is asymptotically AdS, yet the bubble evolution
is Milne-like throughout. Similar to topological NUT Spacetimes, the
presence of a NUT charge induces an ergoregion in the static case. In
section 7 we search for well-behaved bubble solutions with NUT charge in
more than four dimensions. Our search proves fruitless: we find in more than
four dimensions that double (or multiple) analytic continuations give us
spacetimes with at least two timelike coordinates. We then study the four
dimensional Kerr-NUT spacetimes \cite{Mannnutrot} in section 8 and find that
their asymptotic behaviour at late times is similar to that of the zero-NUT
charge case \cite{Aha}, and so such spacetimes can be candidate
time-dependent backgrounds for string theory.\ However the presence of a NUT
charge can induce quantitatively different (and sometimes even asymmetric)
particle creation and temporal evolution of the bubble. \ In section 9 we
consider the AdS generalizations of such spacetimes and find that they are
beset with the same quantum instability problems as their non-rotating
counterparts. \ We conclude with some brief comments about generalizations
to de Sitter spacetimes.

\section{Bubbles from Spherically Symmetric Spacetimes}

The general form of a four-dimensional static, spherically symmetric metric
can be written as 
\begin{equation}
ds^{2}=-F(r)dt^{2}+\frac{dr^{2}}{G(r)}+r^{2}\left( d\theta ^{2}+\sin
^{2}\theta d\phi ^{2}\right)   \label{sphmet}
\end{equation}
where $F(r)$ and $G(r)$ are functions that are determined by the Einstein
equations. \ A bubble solution is obtained by a double analytic continuation
in the time coordinate ($t\rightarrow i\chi $) and in some other combination
of coordinates on the $S^{2}$-section, effectively changing it into a
two-dimensional de Sitter spacetime.\ Three distinct possibilities arise
from this procedure \ 
\begin{align}
ds^{2}& =F(r)d\chi ^{2}+\frac{dr^{2}}{G(r)}+r^{2}\left( -dt^{2}+\cosh
^{2}t\,d\phi ^{2}\right)   \label{bubds1} \\
ds^{2}& =F(r)d\chi ^{2}+\frac{dr^{2}}{G(r)}+r^{2}\left( -dT^{2}+\sinh
^{2}T\,d\Phi ^{2}\right)   \label{bubds2} \\
ds^{2}& =F(r)d\chi ^{2}+\frac{dr^{2}}{G(r)}+r^{2}\left( -d{\cal T}^{2}+\exp
\left( 2{\cal T}\right) \,d{\cal \varphi }^{2}\right)   \label{bubds3}
\end{align}
that are characterized by the topology of the two-dimensional de Sitter
spacetime at every fixed $\left( r,\chi \right) $. \ Locally these are all
equivalent, being related to one another by the coordinate transformations 
\begin{eqnarray}
&&\sinh t=\sinh T\cosh \Phi =\left( 1+\frac{{\cal \varphi }^{2}}{2}\right)
\sinh {\cal T}+\frac{{\cal \varphi }^{2}}{2}\cosh {\cal T}  \label{desittfmV}
\\
&&\cosh t\cos \phi =\cosh T={\cal \varphi }\exp {\cal T}  \label{desittfmX}
\\
&&\cosh t\sin \phi =\sinh T\sinh \Phi =\left( 1-\frac{{\cal \varphi }^{2}}{2}%
\right) \cosh {\cal T}-\frac{{\cal \varphi }^{2}}{2}\sinh {\cal T}
\label{desittfmW}
\end{eqnarray}
Globally they are all equivalent provided the coordinates $\left( \phi ,\Phi
,\varphi \right) $ are all unwrapped. \ However if any or all of them are
identified, then the de Sitter spacetime corresponds three distinct
evolutions for the circle geometry they describe. For the metric (\ref
{bubds1}) the de Sitter evolution is that of a circle with exponentially
large initial radius as $t\rightarrow -\infty $ that exponentially shrinks
with proper time to a minimal value and then expands again \ -- the
worldsheet is like a candlestick. For (\ref{bubds2}) the circle begins with
zero radius at $T=0$ and then expands exponentially like a bowl. For (\ref
{bubds3}) the circle begins with infinitesimally small initial radius at $%
{\cal T}\rightarrow -\infty $ and exponentially expands with increasing
proper time like a trumpet. \ 

The bubble is located at the largest positive root $r_{0}$ of $F(r)$,
assumed to be at least as great as the largest positive root of $G(r)$
(otherwise there is a wormhole geometry). \ The coordinate $\chi $ can
either be unwrapped or identified with period $2\pi /\sqrt{F^{\prime }\left(
r_{0}\right) \sqrt{G\left( r_{0}\right) }}$ unless $G\left( r_{0}\right) =0$%
, in which $\chi $ must have period $4\pi /\sqrt{F^{\prime }\left(
r_{0}\right) G^{\prime }\left( r_{0}\right) }$. \ The radial variable is
then restricted to the range $r\geq r_{0}$ and the solutions (\ref{bubds1}-%
\ref{bubds3}) have a $U(1)\otimes SO\left( 2,1\right) $ symmetry. \ The
spacetime is geodesically complete for each. \ At any fixed time, $r=r_{0}$
corresponds to the circle of minimal circumference, referred to as the
bubble. This circumference exponentially contracts to $t=0$ after which time
it exponentially expands for the solution (\ref{bubds1}). \ For (\ref{bubds2}%
) this circumference has zero size at $T=0$ after which it expands
exponentially. This spacetime could matched smoothly at $T=0$ onto any
spacetime with a hypersurface of matching extrinsic curvature and 3-metric
(including its time-reversed counterpart), and so can be regarded as
describing the nucleation of a bubble (or its collapse if $T\rightarrow -T$%
). \ For (\ref{bubds3}) the circumference exponentially expands from
infinitesimally small values beginning at ${\cal T}\rightarrow -\infty $ .
It is not time-reversal invariant, and so has a counterpart in which the
bubble undergoes exponential contraction.

It is straightforward to show that any null curve in spacetimes for which
the bubble is expanding has $\left| \dot{\Psi}\right| \leq e^{-\tau }\left| 
\dot{\tau}\right| $ at late times, where $\left( \Psi ,\tau \right) $ is any
pair $\left( \phi ,t\right) ,\left( \Phi ,T\right) ,\left( {\cal \varphi },%
{\cal T}\right) $, \ and the overdot refers to a derivative with respect to
proper time. Hence observers at different values of $\Psi $ will eventually
lose causal contact since at late times any null geodesic can only traverse
exponentially vanishing changes in $\Psi $. Null geodesics at fixed $\Psi $
obey the equation 
\begin{equation}
\dot{r}^{2}-\frac{E^{2}}{r^{2}}G+\frac{G}{F}p_{\chi }^{2}=0
\label{nullfixphi}
\end{equation}
where $p_{\chi }=\dot{\chi}F$ is the conserved momentum in the $\chi $%
-direction and $E=r^{2}\dot{\tau}$ is the conserved energy. \ Eq. (\ref
{nullfixphi}) is an equation for a particle in an effective potential $V=$\ $%
\frac{G}{F}p_{\chi }^{2}-\frac{E^{2}}{r^{2}}G$, with all allowed motions
having $V<0$. If $G\left( r_{0}\right) =0$ and the spacetime is
asymptotically flat, then null geodesics oscillate between some minimal and
maximal values of $r$, and so by appropriate choices of \ $\left( E,p_{\chi
}\right) $ observers at any two differing values of $r$ can be causally
connected. However if $G\left( r_{0}\right) \neq 0$ then the effective
potential diverges at $r=r_{0}$, and so there will always be some region $%
r_{\min }>r\geq r_{0}$ which is not in causal contact with observers in
regions $r>r_{\min }$. \ If the spacetime is not asymptotically flat then
the effective potential may become positive for some $r>r_{\max }$, in which
case observers in this region are causally disconnected from the region $%
r<r_{\max }$. Since $\left| \dot{\chi}\right| =\left| p_{\chi }\right|
/F_{\max }$ it is always possible for observers at different values of \ $%
\chi $ to remain in causal contact at fixed $r$, and so any two observers in
causal contact at differing values of $r$ will also be in causal contact if
they are also at differing values of $\chi $.

The above results are generalizable to higher-dimensional spacetimes \cite
{Bir,Bala}. \ The same classical stability and quantum instability
properties hold for all of these bubbles as for the ones analyzed in ref. 
\cite{Aha}.

\section{Taub-Nut-AdS Bubbles}

The (3+1)-dimensional Taub-Nut-AdS solution is \cite{MannNUT,mynut}: 
\begin{equation}
ds^{2}=-F(r)\left( dt+2N\cos \theta d\phi \right) ^{2}+\frac{dr^{2}}{F(r)}%
+(r^{2}+N^{2})\left( d\theta ^{2}+\sin ^{2}\theta d\phi ^{2}\right)
\label{TNLO}
\end{equation}
where 
\begin{equation}
F(r)=\frac{r^{4}+\left( \ell ^{2}+6N^{2}\right) r^{2}-2m\ell
^{2}r-N^{2}\left( \ell ^{2}+3N^{2}\right) }{(r^{2}+N^{2})\ell ^{2}}
\label{FTNLO}
\end{equation}
where $N$ is the nonvanishing Nut charge, $l$ is related to the cosmological
constant $\Lambda <0,$ by the relation $l=\sqrt{\frac{-\Lambda }{3}}$. Its
Euclidean version is found by setting $t\rightarrow i\chi ,N\rightarrow in$: 
\begin{equation}
ds^{2}=+F_{E}(r)\left( d\chi +2n\cos \theta d\phi \right) ^{2}+\frac{dr^{2}}{%
F_{E}(r)}+(r^{2}-n^{2})\left( d\theta ^{2}+\sin ^{2}\theta d\phi ^{2}\right)
\label{TNEU}
\end{equation}
where 
\begin{equation}
F_{E}(r)=\frac{r^{4}+\left( \ell ^{2}-6n^{2}\right) r^{2}-2m\ell
^{2}r+n^{2}\left( \ell ^{2}-3n^{2}\right) }{(r^{2}-n^{2})\ell ^{2}}
\label{FTNEU}
\end{equation}
The coordinate $\chi $ parametrizes a circle, which is fibered over the
sphere parametrized by $(\theta ,\phi ).$ The geometry of a constant-$r$
surface is that of a Hopf fibration of $S^{1}$ over $S^{2}$. To avoid
singularities the periodicity of $\chi $ must simultaneously be $8\pi n$ and 
$4\pi /F_{E}^{\prime }\left( r_{0}\right) $, which forces one of two
conditions: 
\begin{equation}
m=n\left( 1-\frac{4n^{2}}{\ell ^{2}}\right) \text{ if}\ r_{0}=n
\label{nutmass}
\end{equation}
or 
\begin{equation}
m=\frac{r_{b}^{4}+\left( r_{b}^{2}+n^{2}\right) \ell
^{2}-6n^{2}r_{b}^{2}-3n^{4}}{2\ell ^{2}r_{b}}\text{ if}\ r_{0}=r_{b}=\frac{%
\ell ^{2}}{12n}\left( 1\pm \sqrt{1-48\frac{n^{2}}{\ell ^{2}}+144\frac{n^{4}}{%
\ell ^{4}}}\right) \text{\ }  \label{boltmass}
\end{equation}
depending on the value of the largest root $r_{0}$ of $F\left( r\right) $. \
The solution (\ref{nutmass}) is referred to as the NUT solution -- the fixed
point of the $\partial /\partial \chi $ isometry is 0-dimensional. For the
special case $m=0$ and $n=\ell /2$, it is pure AdS. \ Solution (\ref
{boltmass}) is referred to as the bolt solution -- the fixed point of the $%
\partial /\partial \chi $ isometry is 2-dimensional, as is the case for a
black hole. \ 
\begin{figure}[tbp]
\begin{center}
\epsfig{file=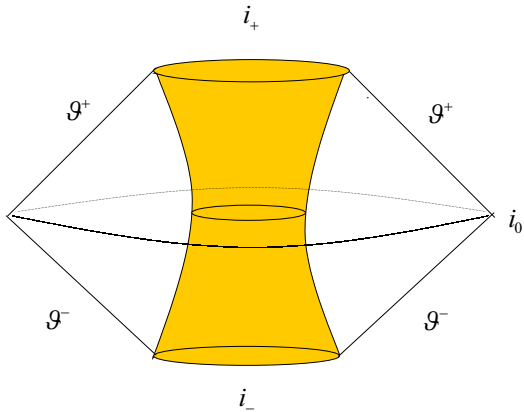,width=0.6\linewidth}
\end{center}
\caption{{}The Penrose diagram of \ an $\ell \rightarrow \infty $ Taub-Nut
bubble obtained from (\ref{tnbub1}) with $\protect\chi $-dependence
suppressed. The space inside the shaded region is absent and the surface of
that region is the bubble.}
\label{fig1}
\end{figure}
We can obtain bubble spacetimes by forming several other analytic
continuations under the constraint that there be only one timelike
direction. \ These metrics are all generalizations of (\ref{bubds1}-\ref
{bubds3}): 
\begin{equation}
ds^{2}=F(r)\left( d\chi +2N\sinh td\phi \right) ^{2}+\frac{dr^{2}}{F(r)}%
+\left( r^{2}+N^{2}\right) \left( -dt^{2}+\cosh ^{2}t\,d\phi ^{2}\right)
\label{tnbub1}
\end{equation}
\begin{equation}
ds^{2}=F(r)\left( d\chi +2N\cosh Td\Phi \right) ^{2}+\frac{dr^{2}}{F(r)}%
+\left( r^{2}+N^{2}\right) \left( -dT^{2}+\sinh ^{2}T\,d\Phi ^{2}\right)
\label{tnbub2}
\end{equation}
\begin{equation}
ds^{2}=F(r)\left( d\chi +2N\exp {\cal T}\,d{\cal \varphi }\right) ^{2}+\frac{%
dr^{2}}{F(r)}+(r^{2}+N^{2})\left( -d{\cal T}^{2}+\exp \left( 2{\cal T}%
\right) \,d{\cal \varphi }^{2}\right)  \label{tnbub3}
\end{equation}
and all are exact solutions to the Einstein equations. \ Each is nonsingular
and geodesically complete. \ For large $r$ they all can easily be shown to
approach AdS$_{4}$ if $\ell \neq 0$. The geometry of sections of fixed $r$
at each constant time slice is that of a\ twisted torus $\widetilde{T^{2}}.$
\ Using the notation of the previous section, at each constant time slice $%
\tau =(t,T,{\cal T}\,)$, the twisted torus has one fundamental circle
respectively parametrized by a coordinate $\Psi =(\phi ,\Phi ,\varphi ).$
The other twisted circle is respectively parameterized by a shifted
coordinate $\Xi =(\chi +2N\phi \sinh t,\chi +2N\Phi \cosh T,\chi +2N{\cal %
\varphi }\exp {\cal T}\,{\cal )}$. The torus becomes exponentially more
twisted with time. If $N=0,$ the geometry of fixed $r$ sections is $%
S^{1}\times $dS$_{2}.$%
\begin{figure}[tbp]
\begin{center}
\epsfig{file=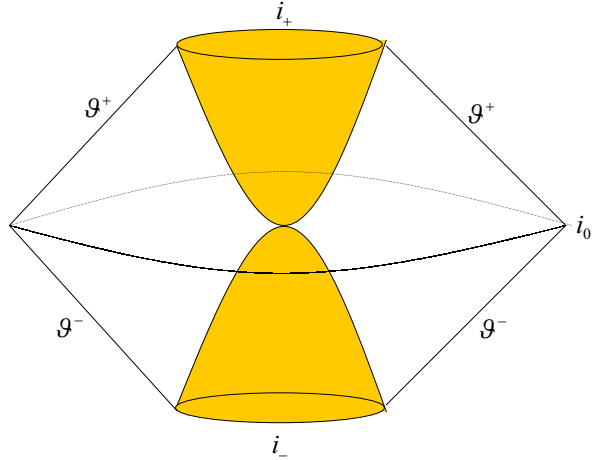,width=0.6\linewidth}
\end{center}
\caption{{}The Penrose diagram of \ an $\ell \rightarrow \infty $ Taub-Nut
bubble obtained from (\ref{tnbub2}) with $\protect\chi $-dependence
suppressed. The space inside the shaded region is absent and the surface of
that region is the bubble.}
\label{fig2}
\end{figure}

\begin{figure}[tbp]
\begin{center}
\epsfig{file=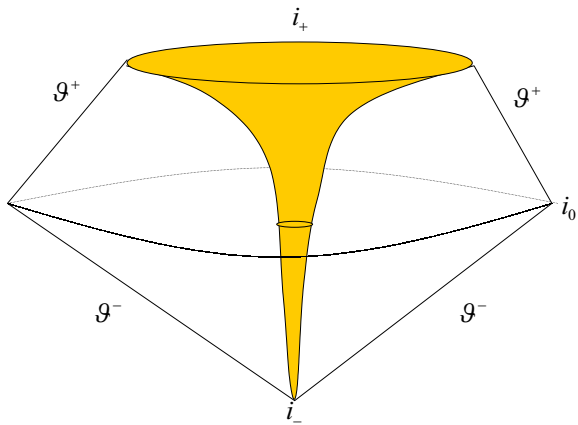,width=0.6\linewidth}
\end{center}
\caption{{}The Penrose diagram of \ an $\ell \rightarrow \infty $ Taub-Nut
bubble obtained from (\ref{tnbub3}) with $\protect\chi $-dependence
suppressed. The space inside the shaded region is absent and the surface of
that region is the bubble.}
\label{fig3}
\end{figure}

The quartic function $F(r)$ in eq. (\ref{FTNLO}) has two real roots, one at $%
r>0$ and one at $r<0$ respectively. \ Without loss of generality we can
locate the bubble at the positive root of $F(r)$, \ since bubble solutions
exist for any values of the parameter $m$, positive, zero or negative
(unlike the spherical bubbles of the previous section). The Penrose diagrams
of the asymptotically flat bubble spacetimes obtained from (\ref{tnbub1})-(%
\ref{tnbub3}) are given in figures (\ref{fig1})-(\ref{fig3}). \ The Penrose
diagrams of the asymptotically AdS bubble spacetimes (\ref{tnbub1})-(\ref
{tnbub3}) are similar to (\ref{fig1})-(\ref{fig3}), by changing the diagonal
and off-diagonal lines in these figures to a cylinder which encloses the
bubble. The bubble is located on the surface of the shaded region at $%
r=r_{0} $, where $r_{0}$\ is the root of $F(r).$%
In the asymptotically flat case ($\ell \rightarrow \infty $) we obtain from (%
\ref{FTNLO}) 
\begin{equation}
r_{0}=m+\sqrt{m^{2}+N^{2}}  \label{bubflatroot}
\end{equation}
for the location of the bubble. If $\ m\geq 0$ the function $F(r)$
monotonically approaches unity as $r\rightarrow \infty $, whereas if $m<0$
it has a local maximum. \ \ In the asymptotically AdS case $F(r)$ has a
local maximum and minimum outside of the bubble if $m\geq 0$; otherwise it
monotonically increases. The only restriction on the periodicity of $\chi $
is that it has the value$\frac{4\pi }{F^{\prime }\left( r_{0}\right) }$
which equals $4\pi \left( \sqrt{m^{2}+N^{2}}+m\right) $ in the
asymptotically flat case. \ 

Using the notation of the previous section, it is straightforward to check
that observers at different values of $\Psi $ eventually fall out of causal
contact, and that observers at fixed $\Psi $ but differing values of $\left(
r,\chi \right) $ can remain in causal contact in both the asymptotically
flat and AdS cases. \ \ 

Consider next the double analytic continuation 
\begin{equation}
(t,\phi )\rightarrow i(\tilde{\chi},\hat{\tau})  \label{a1}
\end{equation}
applied to the metric (\ref{TNLO}): 
\begin{equation}
ds^{2}=+F(r)\left( d\tilde{\chi}+2N\cos \theta d\hat{\tau}\right) ^{2}+\frac{%
dr^{2}}{F(r)}+(r^{2}+N^{2})\left( d\theta ^{2}-\sin ^{2}\theta d\hat{\tau}%
^{2}\right)  \label{TNB1}
\end{equation}
This metric is simply a local patch of one of the solutions (\ref{bubds1}-%
\ref{bubds3}); for example under the coordinate transformation 
\begin{eqnarray}
\cos \theta &=&\cos \phi \cosh t  \nonumber \\
\tanh \hat{\tau} &=&\sin \phi \coth t  \label{statcoordtrans} \\
\tilde{\chi} &=&\chi -2N\left[ \tanh ^{-1}\left( \coth \frac{t}{2}\tan \frac{%
\phi }{2}\right) -\tanh ^{-1}\left( \tanh \frac{t}{2}\tan \frac{\phi }{2}%
\right) \right]  \nonumber
\end{eqnarray}
the metric (\ref{tnbub1}) is recovered. The lapse function is $\sqrt{%
r^{2}+N^{2}}\sin \theta $ and so is non-vanishing everywhere except $\theta
=0,\pi $. \ 

In the coordinates (\ref{TNB1}) the metric has a Killing vector $\xi =\frac{%
\partial }{\partial \hat{\tau}}$ and so describes a static bubble, not
evolving in time. The norm of this Killing vector is 
\begin{equation}
\xi \cdot \xi =4N^{2}F(r)-[r^{2}+N^{2}+4N^{2}F(r)]\sin ^{2}\theta
\label{timecoeff}
\end{equation}
and becomes spacelike unless $\hat{\theta}(r)\leq \left| \theta \right| \leq
\pi -\hat{\theta}(r),$ where 
\begin{equation}
\hat{\theta}(r)=\tan ^{-1}\left[ \frac{2N\sqrt{F(r)}}{\sqrt{r^{2}+N^{2}}}%
\right]  \label{thetalimits}
\end{equation}
\bigskip taking $\theta \in \left( -\pi ,\pi \right) $. These limits define
an `ergocone' for the spacetime, in which $\frac{d\tilde{\chi}}{d\hat{\tau}}%
<0$ ($>0$) for all timelike curves in the region $\theta <\hat{\theta}(r)$ ($%
\theta >\pi -\hat{\theta}(r)$). For $\ell =\infty $ (or $N=0$) the angle (%
\ref{thetalimits}) becomes vanishingly small at large $r$, and the ergocone
shrinks to a pair of lines on the $\theta =0,\pi $ axis where the lapse
function vanishes.\bigskip

Observers using the coordinates (\ref{TNB1}) can, on a surface fixed $r$
(with respective spacelike/timelike unit normals $\hat{n}=\sqrt{F}\partial
/\partial r$ and $\hat{u}=\sin \theta \sqrt{r^{2}+N^{2}}\partial /\partial 
\hat{\tau}$), associate a quasilocal mass density with this spacetime

\begin{equation}
\frac{m}{8\pi }\sin \theta +O(r^{-1})  \label{massdens}
\end{equation}
where we have used the counterterm method \cite{MannNUT,Das} to compute the
mass density. We therefore obtain a total quasilocal mass $M$ of 
\begin{equation}
M=2m\left( \sqrt{m^{2}+N^{2}}+m\right) +O(\varepsilon ^{2})  \label{Mflat}
\end{equation}
where the integration over $\theta $ is from {\rm \ }$-\pi +\varepsilon $%
{\rm \ }to $-\varepsilon $\ and then from $\varepsilon $\ to{\rm \ }$\pi
-\varepsilon .${\bf \ }We note that if $N\neq 0$ the total mass can be
either positive or negative, a situation reminiscent of one that can occur
for topological black holes \cite{Mannblackneg}.\ 

If $\ell $ is finite then this surface penetrates the ergocone, where $%
\theta $ is restricted to the interval $2\frac{N}{\ell }\leq \left| \theta
\right| \leq \pi -2\frac{N}{\ell }$ at large $r$. Since the lapse function
is positive in this region, we can perform a similar calculation; the mass
density is again given by (\ref{massdens}) and we obtain

\begin{equation}
M=\frac{2m\ell ^{3}r_{0}}{\left( \ell ^{2}+3N^{2}+3r_{0}^{2}\right) \sqrt{%
\ell ^{2}+4N^{2}}}+O(\varepsilon ^{2})  \label{Mads}
\end{equation}
where $r_{0}$ is the largest positive root of $F(r)$.

Somewhat counterintuitively, there is no angular momentum associated with
the Killing vector $\frac{\partial }{\partial \tilde{\chi}}$ in either case,
despite the presence of an ergoregion. \ It is straightforward to show that
both the extrinsic curvature and counterterm contributions to the angular
momentum density are identically zero.

Interpretation of the preceding results must be taken with some care. Since
at fixed $\left( r,\hat{\tau}\right) $ the metric (\ref{TNB1}) describes a
torus, observers at $\theta >0$ are separated from observers at $\theta <0$
by a line of vanishing lapse, and so their surface does not actually enclose
the bubble. However observers on one side of this boundary (e.g. $%
\varepsilon <\theta <\pi -\varepsilon $) could infer the above results by
making use of the $\theta \rightarrow -\theta $ symmetry of their metric and
assuming that a corresponding set of observers existed on the other side
that were making similar calculations/measurements.

\section{Boundary Stress Energies}

The large-$r$ boundary stress-energy tensor for the bubble solution (\ref
{tnbub1}) can be calculated by the well known counterterm method, 
\begin{equation}
\begin{array}{c}
8\pi GT_{t}^{t}=\frac{m\ell }{r^{3}}+\frac{1}{4\ell r^{4}}%
(63N^{4}+20N^{2}\ell ^{2}+\ell ^{4})+O(\frac{1}{r^{5}}) \\ 
8\pi GT_{\chi }^{\chi }=-2\frac{m\ell }{r^{3}}-\frac{1}{4\ell r^{4}}%
(105N^{4}+30N^{2}\ell ^{2}+\ell ^{4})+O(\frac{1}{r^{5}}) \\ 
8\pi GT_{\phi }^{\phi }=\frac{m\ell }{r^{3}}+\frac{1}{4\ell r^{4}}%
(63N^{4}+20N^{2}\ell ^{2}+\ell ^{4})+O(\frac{1}{r^{5}}) \\ 
8\pi GT_{\phi }^{\chi }=-N\{6\frac{m\ell }{r^{3}}+\frac{1}{\ell r^{4}}%
(4N^{2}+\ell ^{2})(21N^{2}+\ell ^{2})\}\sinh t+O(\frac{1}{r^{5}})
\end{array}
\label{TNB3STRESS}
\end{equation}
As one can see, $T_{t}^{t}>0,$ which means that the bubble solution has a
negative mass and $T_{\chi }^{\chi }<0,$ which means there exists a negative
pressure in the $\chi $ direction. We see that the presence of a NUT charge
enhances these respective values in the next-to-leading-order term. The
boundary energy momentum tensor components for the bubble solution (\ref
{tnbub2}) also are given by the same relations (\ref{TNB3STRESS}) changing
coordinates to $t\rightarrow T,\phi \rightarrow \Phi $ and $\sinh
t\rightarrow \cosh T$ in the last equation of (\ref{TNB3STRESS}). Similarly,
the components for the bubble solution (\ref{tnbub3}) also are given in (\ref
{TNB3STRESS}) changing coordinates $t\rightarrow {\cal T},\phi \rightarrow 
{\cal \varphi }$ and sinh$t\rightarrow \exp {\cal T}\,$\ in the last
equation.

We can compare the above boundary stress-energy tensor components of the
bubble solution with the components of boundary energy momentum tensor for
the Taub-Nut-AdS spacetime (\ref{TNLO}). The non-vanishing components of the
latter are given by: 
\begin{equation}
\begin{array}{c}
8\pi GT_{t}^{t}=-2\frac{m\ell }{r^{3}}-\frac{1}{4\ell r^{4}}%
(105N^{4}+30N^{2}\ell ^{2}+\ell ^{4})+O(\frac{1}{r^{5}}) \\ 
8\pi GT_{\theta }^{\theta }=\frac{m\ell }{r^{3}}+\frac{1}{4\ell r^{4}}%
(63N^{4}+20N^{2}\ell ^{2}+\ell ^{4})+O(\frac{1}{r^{5}}) \\ 
8\pi GT_{\phi }^{\phi }=\frac{m\ell }{r^{3}}+\frac{1}{4\ell r^{4}}%
(63N^{4}+20N^{2}\ell ^{2}+\ell ^{4})+O(\frac{1}{r^{5}}) \\ 
8\pi GT_{\phi }^{t}=-N\{6\frac{m\ell }{r^{3}}+\frac{1}{\ell r^{4}}%
(4N^{2}+\ell ^{2})(21N^{2}+\ell ^{2})\}\cos \theta +O(\frac{1}{r^{5}})
\end{array}
\label{TNLOSTRESS}
\end{equation}
As a consistency check, we see that in special case of $N=0,$ for which (\ref
{TNLO}) reduces to the 4-dimensional Schwarzschild-AdS black hole, the
covariant energy-momentum tensor components of \ Taub-Nut-AdS spacetime (\ref
{TNLO}) reduce to the same as obtained in \cite{Bala2}.

The equivalence of the non-vanishing boundary energy-momentum tensor of \
the bubble solution with the components of boundary energy momentum tensor
for the Taub-Nut-AdS spacetime (\ref{TNLO}) is a direct consequence of
analytic continuation. So the boundary CFT quantities related to the
energy-momentum tensor (like the central charge or Casimir energy) are the
same for the two nutty bubble spacetimes and the Taub-Nut-AdS spacetime.
Such an equivalence was observed for the boundary energy-momentum tensors of
\ bubble solutions derived from analytic continuation of the
five-dimensional Schwarzschild-AdS metric and Schwarzschild-AdS metric \cite
{Bala}. Although the boundary theory is not known for the bubble metrics (%
\ref{tnbub1})-(\ref{tnbub3}) (in contrast to the 5d-Schwarzschild bubble 
\cite{Bala}, for which the boundary theory is ${\cal N}=4$ SYM living on $%
S^{1}\times $dS$_{3}$), since the boundary is conformally flat we can use
the following relation for the stress tensor \cite{Birrel}: 
\begin{equation}
\left\langle T_{\mu }^{\nu }\right\rangle =-\frac{1}{16\pi ^{2}}%
(A^{(1)}H_{\mu }^{\nu }+B^{(3)}H_{\mu }^{\nu })+\widetilde{T_{\mu }^{\nu }}
\label{s}
\end{equation}
where $^{(1)}H_{\mu }^{\nu \text{ \ }}$and $^{(3)}H_{\mu }^{\nu }$ are
conserved quantities constructed from the curvatures via 
\begin{equation}
\begin{array}{c}
^{(1)}H_{\mu \nu }=2R_{;\mu \nu }-2g_{\mu \nu }\square R-\frac{1}{2}g_{\mu
\nu }R^{2}+2RR_{\mu \nu } \\ 
^{(3)}H_{\mu \nu }=\frac{1}{12}R^{2}g_{\mu \nu }-R^{\rho \sigma }R_{\rho \mu
\sigma \nu }
\end{array}
\label{H13}
\end{equation}
and $\widetilde{T_{\mu }^{\nu }}$ is the traceless part. For the boundary
metric of the bubble (\ref{tnbub1}), the functions $^{(1)}H_{\mu }^{\nu 
\text{ \ }}$and $^{(3)}H_{\mu }^{\nu }$ are: 
\begin{equation}
\begin{array}{c}
^{(1)}H_{\chi }^{\chi \text{ \ }}=-\frac{1}{2}\frac{9N^{4}+10N^{2}\ell
^{2}+\ell ^{4}}{\ell ^{8}} \\ 
^{(1)}H_{t}^{t\text{ \ }}=^{(1)}H_{\varphi }^{\varphi \text{ \ }}=\frac{%
7N^{4}+10N^{2}\ell ^{2}+3\ell ^{4}}{2\ell ^{8}}
\end{array}
\label{H1}
\end{equation}
\begin{equation}
\begin{array}{c}
^{(3)}H_{\chi }^{\chi \text{ \ }}=-\frac{1}{12}\frac{5N^{4}+26N^{2}\ell
^{2}+9\ell ^{4}}{\ell ^{8}} \\ 
^{(3)}H_{t}^{t\text{ \ }}=^{(3)}H_{\varphi }^{\varphi \text{ \ }}=-\frac{1}{%
12}\frac{37N^{4}+18N^{2}\ell ^{2}+5\ell ^{4}}{\ell ^{8}}
\end{array}
\label{H3}
\end{equation}
By fixing the constants $A$ and $B$ from the boundary theory, the result of (%
\ref{s}) is in agreement with (\ref{TNB3STRESS}) in the limit $r\rightarrow
\infty $, since the two constants must vanish due to absence of the
conformal anomaly in the odd dimensional boundary theory. Consequently the
above comparison of \ the stress tensor (\ref{TNB3STRESS}) to (\ref{s}) does
not result in any non-trivial connection between them. In other words, in
the limit $r\rightarrow \infty $\ , all of the stress tensor components (\ref
{TNB3STRESS}) vanish and since the boundary conformal anomaly vanishes in
odd dimensions, so the boundary stress tensor (\ref{s}) also vanishes.

\section{Topologically Nutty Bubbles}

These bubbles have in four-dimensional Euclidean space the general form 
\begin{equation}
ds^{2}=F_{T}(r)\left( d\chi +2N\cosh \theta d\phi \right) ^{2}+\frac{dr^{2}}{%
F_{T}(r)}+(r^{2}+N^{2})\left( d\theta ^{2}+\sinh ^{2}\theta d\phi ^{2}\right)
\label{topnut}
\end{equation}
where 
\begin{equation}
F_{T}(r)=\frac{r^{4}+\left( -\ell ^{2}+6N^{2}\right) r^{2}-2m\ell
^{2}r-N^{2}\left( -\ell ^{2}+3N^{2}\right) }{(r^{2}+N^{2})\ell ^{2}}
\label{Ftopnut}
\end{equation}
There are three distinct analytic continuations: 
\begin{equation}
ds^{2}=F_{T}(r)\left( d\chi +2N\cosh \theta dt\right) ^{2}+\frac{dr^{2}}{%
F_{T}(r)}+(r^{2}+N^{2})\left( d\theta ^{2}-\sinh ^{2}\theta dt^{2}\right)
\label{top1}
\end{equation}
\begin{equation}
ds^{2}=F_{T}(r)\left( d\chi +2N\sinh \theta dt\right) ^{2}+\frac{dr^{2}}{%
F_{T}(r)}+(r^{2}+N^{2})\left( d\theta ^{2}-\cosh ^{2}\theta dt^{2}\right)
\label{top2}
\end{equation}
\begin{equation}
ds^{2}=F_{T}(r)\left( d\chi +2Ne^{\theta }dt\right) ^{2}+\frac{dr^{2}}{%
F_{T}(r)}+(r^{2}+N^{2})\left( d\theta ^{2}-e^{2\theta }dt^{2}\right)
\label{top3}
\end{equation}
and these all satisfy the Einstein equations. \ 

The $\theta $-coordinate is no longer periodic and takes on all real values.
These metrics can, after a coordinate transformation be written in the form 
\begin{equation}
ds^{2}=F_{T}(r)\left( d\chi +2Nxdt\right) ^{2}+\frac{dr^{2}}{F_{T}(r)}%
+(r^{2}+N^{2})\left( \frac{dx^{2}}{x^{2}+k}-\left( x^{2}+k\right)
dt^{2}\right)  \label{topxk}
\end{equation}
where $k=\left( -1,1,0\right) $ for the respective cases (\ref{top1}-\ref
{top3}). \ They are all locally equivalent to each other under a coordinate
transformation, and so it is tempting to conclude that they are globally
equivalent since the $t$-coordinate is not periodic. \ However at fixed $%
\left( \chi ,r\right) $ the $\left( x,t\right) $ section describes (for $N=0$%
) a two-dimensional anti de Sitter spacetime. \ This spacetime can have
non-trivial identifications \cite{robbisrael} and so can describe either a
two-dimensional black hole \cite{danrobb} as in eqs. (\ref{top1},\ref{top3})
(the latter being a special case of a massless extremal black hole) or else
standard AdS$_{2}$, as in eq. (\ref{top2}). \ 

Another clear difference between this case and the previous cases is that
the quartic function $F_{T}\left( r\right) $ in (\ref{Ftopnut}) can have
four real roots, which we label $\left( r_{--},r_{-},r_{+},r_{++}\right) $
in increasing order of size. From this perspective these metrics can
represent two different situations.

\begin{figure}[tbp]
\begin{center}
\epsfig{file=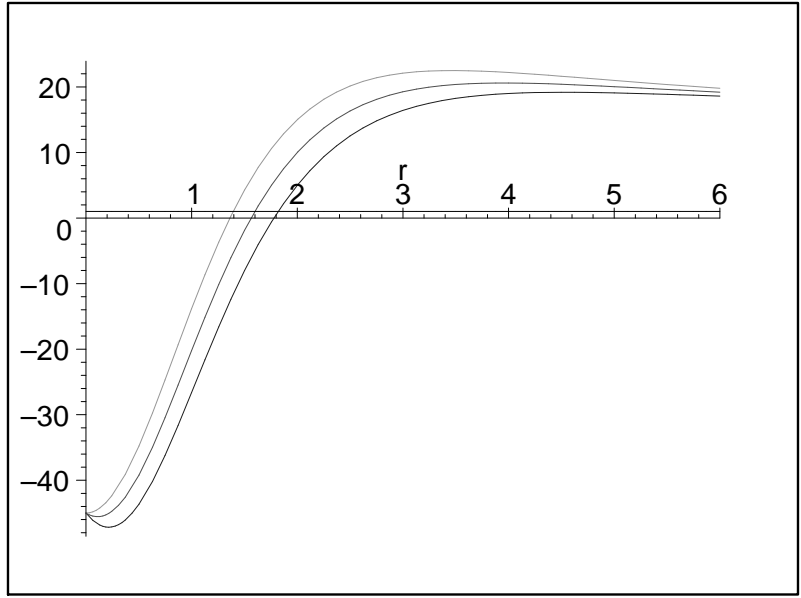,width=0.6\linewidth}
\end{center}
\caption{{}Dependence of the ergoregion on increasing mass parameter. From
left to right the curves are $m=0,5,10$ for $N=2$. The ergoregion is for all
values of $r$ to the right of the intersection point of the curve with the
straight line. The remaining portion of the spacetime is between this
intersection point and where the curve crosses the horizontal axis.}
\label{fig31}
\end{figure}

First, for $r>r_{++}$ \ (or $r<r_{--}$), these spacetime represent static
bubbles. The coordinate $r$ has the range $r_{++}<r<\infty $ (or
alternatively $r_{--}>r>-\infty $), and the $\theta $-coordinate takes on
all possible values. \ At large $\left( r,\theta \right) $ it is
straightforward to show that these metrics are asymptotically AdS. \ Both
invariants $R_{\mu \nu \rho \lambda }R^{\mu \nu \rho \lambda }$ and $\sqrt{-g%
}R_{\mu \nu }^{\quad \,\alpha \beta }\epsilon _{\alpha \beta \rho \lambda
}R^{\mu \nu \rho \lambda }$ are finite everywhere and the spacetime has no
singularities.{\rm \ }Since $x$ is not periodic, the bubble does not close
back on itself -- in this sense it is not really a bubble, but it should be
a valid background for string theory. \ \ In this case each of the metrics
contains an ergoregion given by the expression 
\begin{equation}
\left( \frac{4N^{2}F_{T}\left( r\right) }{r^{2}+N^{2}}-1\right) x^{2}>k
\label{ergotop}
\end{equation}
\ The structure of the ergoregion depends upon the value of $N/\ell $ and $k$%
. Consider first the metric (\ref{top3}), ($k=0$) for which the ergoregion
is independent of $x$. If $N>\ell /2$, then there will be an ergoregion
everywhere except for a very narrow strip of infinite length near the bubble
where $r_{0}<r<r_{c}$, with $4N^{2}F\left( r_{c}\right) =r_{c}^{2}+N^{2}$.
The strip slightly broadens as $m$ increases, as can be seen from figure (%
\ref{fig31}), where the function $\frac{4N^{2}F_{T}\left( r\right) }{%
r^{2}+N^{2}}$versus $r$ is plotted. {\rm \ } {\rm \ } 

As $N\rightarrow \ell /2$ from above $r_{c}$ gets larger, diverging at $%
N=\ell /2$, at which the ergoregion vanishes. Once $N<\ell /2$ there is no
ergoregion.

\begin{figure}[tbp]
\begin{center}
\epsfig{file=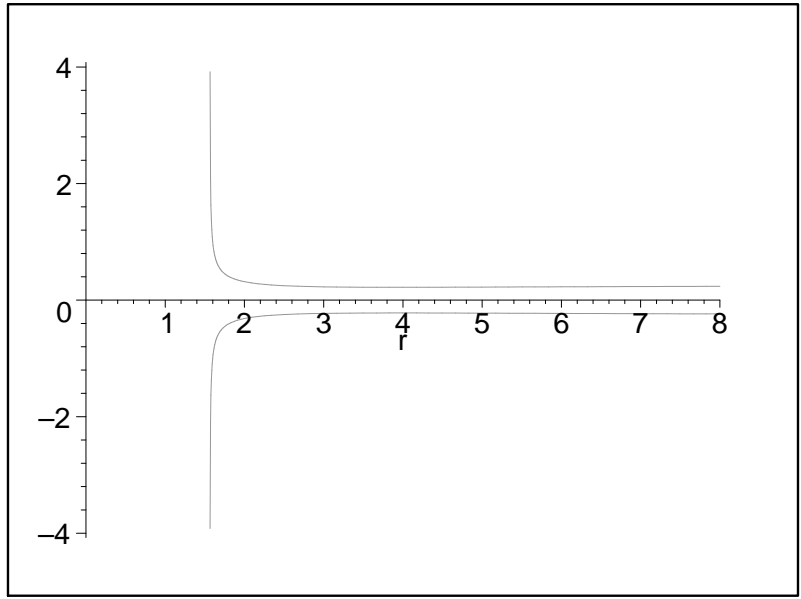,width=0.6\linewidth}
\end{center}
\caption{{}Ergoregion of a topological metric with $k=1$ in the ($r,x$)
plane for $N=2$. The ergoregion is confined to the right upper and lower
halves of plane ($r,x$) bounded by the curves.}
\label{fig32}
\end{figure}

For the metric (\ref{top2}), with $k=1$, there will be no ergoregion for $%
N\leq \ell /2$ . For $N>$ $\ell /2$, the ergoregion is confined to the right
half of the $\left( r,x\right) $ plane bounded by a curve with a vertical
asymptote at $r=r_{c}$, and a horizontal asymptote at $x=1/\sqrt{4N^{2}/\ell
^{2}-1}$. \ This curve has a mirror image about $x=0$ and the ergoregion for 
$x>0$ has a mirror image for negative $x$. A typical ergoregion is plotted
in figure (\ref{fig32}) for special values of $\ell =1,m=5,N=2.$ In this
figure the vertical axis is $x$ and horizontal axis is $r.$ \ 

\begin{figure}[tbp]
\begin{center}
\epsfig{file=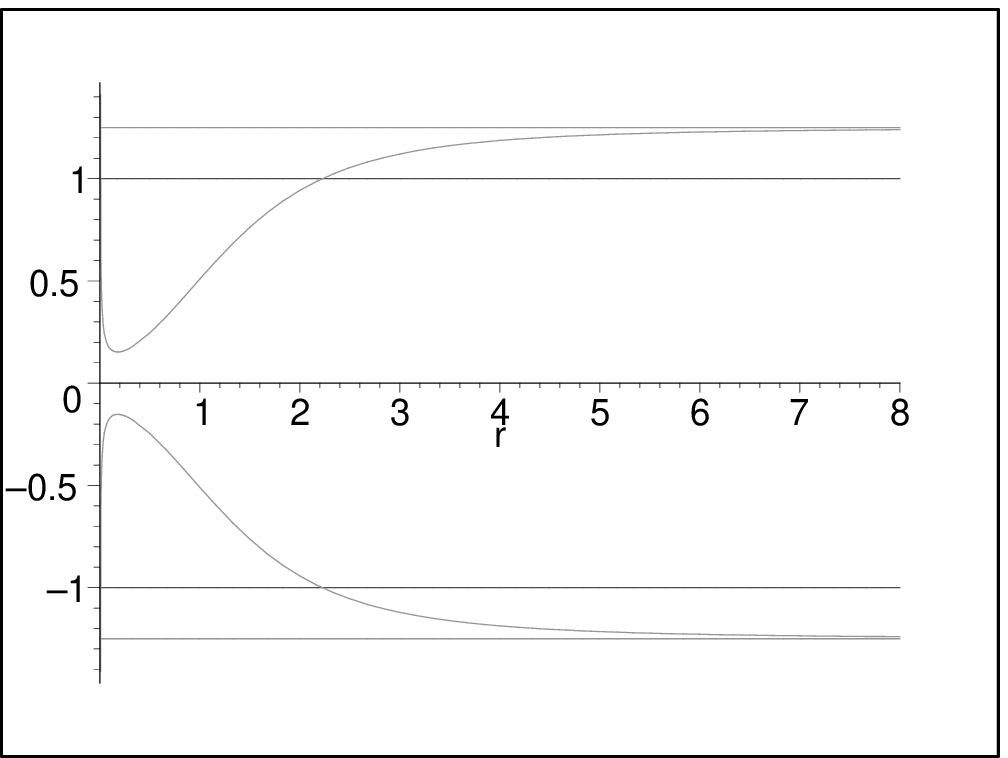,width=0.6\linewidth}
\end{center}
\caption{{}Ergoregion of topological metric with $k=-1$ in plane ($r,x$),
with $N=3\ell /10$. The ergoregion is confined between the horizon locations
at $\left| x\right| =1$ and the curves. The other two horizontal lines are
asymptotes at $\left| x\right| =1/\protect\sqrt{1-4N^{2}/\ell ^{2}}.$}
\label{fig33}
\end{figure}

For the metric (\ref{top1}), with $k=-1$, the ergoregion is confined to a
strip outside the event horizon at $\left| x\right| =1$ bounded by a curve
and its mirror image in $x$ that asymptote for large $r$ at $\left| x\right|
=1/\sqrt{1-4N^{2}/\ell ^{2}}$ for $N<\ell /2$. This curve smoothly
intersects the location of the bubble at the event horizon. A typical
ergoregion is plotted in figure (\ref{fig33}) for special values of $\ell
=1,m=5,N=.3.$ 

\ As $N\rightarrow \ell /2$ from below, the strip becomes increasingly wide,
and at $N=\ell /2$ the boundary curve asymptotes to $x^{3/2}/\sqrt{2m}$. \
For $N>$ $\ell /2$, the boundary of the ergoregion has a vertical asymptote
at $r=r_{c}$, and covers the remainder of the $\left( r,x\right) $ plane for
larger values of $r$.

If $F_{T}(r)$ has 4 real roots, \ a second possibility emerges: the
coordinate $r$ can have the range $r_{-}<r<r_{+}$, and can be transformed to
a periodic variable. \ Removal of conical singularities at $r=r_{\pm }$ is
satisfied by requiring $\left| F_{T}^{\prime }(r_{+})\right| =\left|
F_{T}^{\prime }(r_{-})\right| $, which is 
\begin{equation}
\left( r_{++}-r_{+}\right) \left( r_{+}-r_{--}\right) \left(
r_{-}^{2}+N^{2}\right) =\left( r_{++}-r_{-}\right) \left(
r_{-}-r_{--}\right) \left( r_{+}^{2}+N^{2}\right)  \label{toproots}
\end{equation}
or alternatively, if $\left| r_{-}\right| \neq \left| r_{+}\right| $ 
\begin{equation}
N^{2}=-\left[ r_{++}\left( r_{++}+r_{+}\right) +r_{-}\left(
r_{++}-r_{+}\right) \right]  \label{toproots2}
\end{equation}
since for the function $F_{T}(r)$ we must have $r_{--}=-\left(
r_{++}+r_{+}+r_{-}\right) $. Eq. (\ref{toproots2}) has no solutions since
its right hand side is always negative. \ However if $\left| r_{-}\right|
=\left| r_{+}\right| $ then conical singularities can be removed. \ If $%
r_{-}=-r_{+}$ then $m=0$, whereas if $r_{-}=r_{+}=r_{0}$ we can attempt to
obtain a new spacetime by setting $r_{\pm }=r_{0}\pm \varepsilon $, $%
r=r_{0}+\varepsilon \cos \alpha $%
\begin{equation}
\begin{array}{c}
ds^{2}=\frac{\left( r_{++}-r_{0}\right) \left( r_{0}-r_{--}\right) }{%
(r_{0}^{2}+N^{2})\ell ^{2}}\sin ^{2}\alpha \left( \varepsilon d\chi
+2N\varepsilon xdt\right) ^{2}+\ell ^{2}\frac{d\alpha ^{2}(r_{0}^{2}+N^{2})}{%
\left( r_{++}-r_{0}\right) \left( r_{0}-r_{--}\right) } \\ 
+(r_{0}^{2}+N^{2})\left( \frac{dx^{2}}{x^{2}+k}-\left( x^{2}+k\right)
dt^{2}\right)
\end{array}
\label{topresult}
\end{equation}
and then taking the limit as $\varepsilon \rightarrow 0$. \ This is possible
only if $N=0$, since a rescaling of the coordinates $\left( \chi ,t\right) $
by factors of $\varepsilon ^{-1}$ yields a singularity in the metric.
However if $N=0$ one of the roots becomes $r=0$ at which a singularity is
located.\ 

If $m=0$ then $N<\ell \sqrt{3}/6$ in order that $F_{T}(r)$ be real. \ The
metric can be written as{\bf \ } 
\begin{equation}
\begin{array}{c}
ds^{2}=\frac{{\frak r}_{-}^{2}}{\ell ^{2}}\frac{{\frak r}_{+}^{2}-{\frak r}%
_{-}^{2}\cos ^{2}\alpha }{N^{2}-{\frak r}_{-}^{2}\cos ^{2}\alpha }\sin
^{2}\alpha \left( d\chi +2Nxdt\right) ^{2}+\ell ^{2}d\alpha ^{2}\frac{N^{2}+%
{\frak r}_{-}^{2}\cos ^{2}\alpha }{{\frak r}_{+}^{2}-{\frak r}_{-}^{2}\cos
^{2}\alpha } \\ 
+({\frak r}_{-}^{2}\cos ^{2}\alpha +N^{2})\left( \frac{dx^{2}}{x^{2}+k}%
-\left( x^{2}+k\right) dt^{2}\right)
\end{array}
\label{topresult2}
\end{equation}
where $\alpha \in \left[ 0,\pi \right] $ , $\chi $ has the period $\frac{%
2\pi }{{\frak r}_{-}}\frac{{\frak r}_{-}^{2}+N^{2}}{\left( {\frak r}_{+}^{2}-%
{\frak r}_{-}^{2}\right) }$, and 
\begin{equation}
{\frak r}_{\pm }^{2}=\frac{\ell ^{2}}{2}-3N^{2}\pm \frac{1}{2}\sqrt{%
48N^{4}-16N^{2}\ell ^{2}+\ell ^{4}}  \label{toprootssph}
\end{equation}
Strictly speaking this is not a bubble spacetime: instead the $\left( \alpha
,\chi \right) $ section is the surface of a deformed 2-sphere. \ The
ergoregion is now located in a band about the equator at $\alpha =\frac{\pi 
}{2}$, whose specific boundary curves are complicated functions of $N/\ell $%
. \ If $k=0$ this band is very narrow and exists only for a very small range
of $N$, and it remains constant everywhere. However if $k\neq 0$ the band
will change shape as $x$ increases. \ Note that if we set $\ell \rightarrow
\infty $ then we must have at least one conical singularity, since the
condition \ $\left| F_{T}^{\prime }(r_{+})\right| =\left| F_{T}^{\prime
}(r_{-})\right| $ can't be satisfied unless $N=0$, in which case there is
again a singularity. The black hole ($k=-1$) becomes a compact region,
surrounded by a deformed `ergosphere' for small values of $x$. \ 

Finally, we can continue to the metric 
\begin{equation}
ds^{2}=F_{T}(r)\left( d\chi +2N\cos t\,d\phi \right) ^{2}+\frac{dr^{2}}{%
F_{T}(r)}+(r^{2}+N^{2})\left( -dt^{2}+\sin ^{2}t\,d\phi ^{2}\right)
\label{topnutperiodic}
\end{equation}
which locally can be transformed to one of the metrics (\ref{top1}-\ref{top3}%
) via a change of coordinates; it can be regarded as the extension of the $%
k=-1$ case if $\phi $ is not periodic. \ However if the $\phi $-coordinate
in (\ref{topnutperiodic}) is identified then these metrics are not globally
equivalent, yielding new bubble solutions.

If $\ r>r_{++}$ \ (or $r<r_{--}$), the metric (\ref{topnutperiodic})
describe a bubble spacetime in which the bubble expands from zero size to
the finite size $(r_{++}^{2}+N^{2})$ and then contracts back to zero size
again. All spacetime events are causally connected to one another, and near
the initial expansion (or final contraction) the scale factor is linear in
time and the spacetime behaves like a Milne universe. \ 

If $r_{-}<r<r_{+}$ then the metric (\ref{topnutperiodic}) describes a
compact time-dependent 3-geometry of a twisted circle over a 2-sphere. From
the previous analysis, conical singularities are avoided only if either $m=0$
or $N=0$. In each case the spacetime oscillates, with the $S^{1}$
represented by the $\phi $ coordinate repeatedly expanding and contracting.
There are again no event horizons. These spacetimes are not really bubbles
because the $\left( r,\chi \right) $\ section is a compact space. However
they are candidate time-dependent backgrounds. Since all of the topological
backgrounds (\ref{top1})-(\ref{top3}) and (\ref{topnutperiodic}) are\ four
dimensional they are stable, similar to the other nutty bubbles (\ref{tnbub1}%
)-(\ref{tnbub3}).

We close this section by noticing that in the case of\ (\ref{topnutperiodic}%
), null geodesics relevant for s-waves behaviour satisfy: 
\begin{equation}
\frac{dr^{2}}{(r^{2}+N^{2})F_{T}(r)}=dt^{2}  \label{nulltopgeo}
\end{equation}
The integration involves the square root of a quartic function and so
finding a closed-form relation between coordinates $r$\ and $t$\ for a null
geodesic is quite problematic. The relation between these two coordinates is
essential to obtain a relation between phase differences of \ two geodesics
on $\vartheta ^{+}$ and $\vartheta ^{-}$. This relation can be used as an
evidence for particle creation. We shall not consider this complicated
problem here.\ However in the case of Kerr-Nut bubbles, the integration can
be done and a simple relation is obtained for null geodesics, eq. (\ref
{trnull}), as we shall later consider.

\section{Nutty Rindler Bubbles}

Another class of NUT-charged spacetimes has the metric 
\begin{equation}
ds^{2}=-F_{R}(r)\left( dt+N\upsilon ^{2}d\phi \right) ^{2}+\frac{dr^{2}}{%
F_{R}(r)}+(r^{2}+N^{2})\left( d\upsilon ^{2}+\upsilon ^{2}d\phi ^{2}\right)
\label{lorflat}
\end{equation}
where 
\begin{equation}
F_{R}(r)=\frac{r^{4}+6N^{2}r^{2}-2m\ell ^{2}r-3N^{4}}{(r^{2}+N^{2})\ell ^{2}}
\label{Frind}
\end{equation}
has only two real roots, and the $\left( \upsilon ,\phi \right) $ section
describes a flat two-dimensional space, possibly with identifications. \ 

There are two distinct double analytic continuations: 
\begin{equation}
ds^{2}=F_{R}(r)\left( d\chi +N{\sf t}^{2}d\phi \right) ^{2}+\frac{dr^{2}}{%
F_{R}(r)}+k(r^{2}+N^{2})\left( -d{\sf t}^{2}+{\sf t}^{2}d\phi ^{2}\right)
\label{nutrind}
\end{equation}
where $k=\pm 1$. The $\left( {\sf t},\phi \right) $ section is now a
two-dimensional Milne/Rindler spacetime, which can also be written as 
\begin{equation}
ds^{2}=F_{R}(r)\left( d\chi +N\left( xdy-ydx\right) \right) ^{2}+\frac{dr^{2}%
}{F_{R}(r)}+k(r^{2}+N^{2})\left( dx^{2}-dy^{2}\right)  \label{nutsoliton}
\end{equation}
which is a NUT-charged generalization of the AdS soliton \cite{Adssoliton};
we shall continue to refer to it as a bubble. Without loss of generality the
bubble (\ref{nutsoliton}) can be located at the positive root of $F_{R}(r)$
since $m$ can take on all real values. The cases $k=\pm 1$ are locally
equivalent under a coordinate transformation, but will be globally distinct
if $\phi $ is periodically identified.

When $k=+1$ the bubble undergoes a Milne-type evolution. At any fixed ${\sf t%
}$ the geometry of the bubble is that of a twisted torus, with the
coordinate $\phi $ parametrizing one circle, and with the other twisted
circle parameterized by the shifted coordinate $\Xi =\chi +N{\sf t}\phi $. \ 
{\bf \ }

\begin{figure}[tbp]
\begin{center}
\epsfig{file=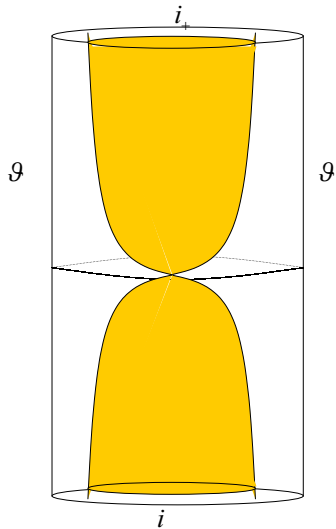,width=0.6\linewidth}
\end{center}
\caption{{}The Penrose diagram of the Rindler bubble (\ref{nutrind}) with $%
k=1.$ $\protect\chi $-dependence suppressed. The space inside the shaded
region is absent and the surface of that region is the bubble. }
\label{fig34}
\end{figure}

Turning next to the causal structure of the bubble (\ref{nutrind}), we
consider two observers at different coordinates $\phi .$\ In the background
of bubble (\ref{nutrind}), a null curve satisfies 
\begin{equation}
F_{R}(r)(\stackrel{\cdot }{\chi }+N{\sf t}^{2}\stackrel{\cdot }{\phi })^{2}+%
\frac{\stackrel{\cdot }{r}^{2}}{F_{R}(r)}+(r^{2}+N^{2})(-\stackrel{\cdot }{%
{\sf t}}^{2}+{\sf t}^{2}\stackrel{\cdot }{\phi }^{2})=0  \label{nullrin}
\end{equation}
\ and so 
\begin{equation}
{\sf t}^{2}\stackrel{\cdot }{\phi }^{2}\leq \text{ }\stackrel{\cdot }{{\sf t}%
}^{2}\Rightarrow \phi \backsim \ln {\sf t}  \label{res2}
\end{equation}
As ${\sf t}\rightarrow \pm {\sf T}$ , we have $\Delta \phi \leq \ln {\sf T}$%
, and so observers at two points with different $\phi $\ can communicate to
each other for arbitrarily large ${\sf T}$. The Penrose diagram of the
bubble spacetime (\ref{nutrind}) with $k=+1$ is given in figure (\ref{fig34}%
). Any radial null ray leaving the surface of the bubble reaches infinity in
a finite amount of coordinate time ${\sf t}$, and there is always an
infinite amount of proper distance between the bubble and spatial infinity.\
The bubble is located on the surface of the shaded region at $r=r_{0}$,
where $r_{0}$\ is the root of $F_{R}(r).$%

When $k=-1$ the geometry is that of a static bubble with an ergoregion.
Writing the metric (\ref{nutrind}) for this case as 
\begin{equation}
ds^{2}=F_{R}(r)\left( d\chi +Nx^{2}d{\sf t}\right) ^{2}+\frac{dr^{2}}{%
F_{R}(r)}+(r^{2}+N^{2})\left( dx^{2}-x^{2}d{\sf t}^{2}\right)
\label{nutrinderg}
\end{equation}
we see that the lapse function is $(r^{2}+N^{2})x^{2}$ and so vanishes at
the Rindler horizon $x=0$. The ergoregion is described by 
\begin{equation}
x^{2}>\frac{r^{2}+N^{2}}{N^{2}F_{R}\left( r\right) }  \label{ergorind}
\end{equation}
and at large $r$ includes the entire $\left( r,x\right) $ plane except for a
strip bounded by a pair of curves opposite the $r$-axis that asymptote to $%
\pm \ell /N$. \ As $r$ decreases, this strip widens, becoming infinitely
wide at the bubble edge. The ergoregion is similar to that given in fig. (%
\ref{fig32}), except that the vertical asymptotes are at the bubble edge $%
r=r_{0}$.

Using the same coordinates as (\ref{nutrinderg}), we have 
\begin{equation}
\stackrel{\cdot }{x}^{2}\leq x^{2}\text{ }\stackrel{\cdot }{{\sf t}}%
^{2}\Rightarrow x\backsim e^{\left| {\sf t}\right| }  \label{res1}
\end{equation}
and so there is no restriction as to the maximum change of coordinate $x$
for points on the null curve as ${\sf t}\rightarrow \pm \infty $. So
observers at any two points with different values of $x$\ can communicate
with each other.

\section{Higher dimensional Taub-Nut-AdS Bubbles?}

In this section we consider the possibility of extending our NUT-charged
bubble solutions to higher dimensions. The six dimensional Taub-Nut-AdS is a
fibration over $S^{2}\times S^{2}$ \cite{awad}: 
\begin{equation}
\begin{array}{c}
ds^{2}=V(r)\left( d\chi +2n\cos \theta _{1}d\phi _{1}+2n\cos \theta
_{2}d\phi _{2}\right) ^{2}+\frac{dr^{2}}{V(r)}+(r^{2}-n^{2}) \\ 
\times \left( d\theta _{1}^{2}+\sin ^{2}\theta _{1}d\phi _{1}^{2}+d\theta
_{2}^{2}+\sin ^{2}\theta _{2}d\phi _{2}^{2}\right)
\end{array}
\label{TNEUC}
\end{equation}
If $r=n$ is the largest root of $V(r)$, then singularities in (\ref{TNEUC})
are avoided by choosing the mass parameter and the function $V(r)$ by

\begin{equation}
m=\frac{4n^{3}(6n^{2}-\ell ^{2})}{3\ell ^{2}}  \label{mEUC}
\end{equation}
and

\begin{equation}
V(r)=\frac{(r-n)(3r^{3}+9nr^{2}+(\ell ^{2}+3n^{2})r+3n(\ell ^{2}-5n^{2}))}{%
3(r+n)^{2}\ell ^{2}}  \label{VTNEUC}
\end{equation}
The coordinate $\tau $ must be periodic with the period $12\pi n.$

In performing a single analytic continuation from the Euclidean metric (\ref
{TNEUC}) we are forced to introduce two timelike directions unless we
analytically continue $\chi $. \ The reason is that there is a single NUT
charge $n$ and all analytic continuations in one of the $S^{2}$ sections
forces $n\rightarrow iN$. \ This in turn forces an additional analytic
continuation in the other $S^{2}$ section, yielding two timelike directions.
\ 

For example performing the following analytic continuation 
\begin{equation}
(\theta _{1},\phi _{2},n)\rightarrow (i\theta _{1}+\frac{\pi }{2},-i\phi
_{2},iN)  \label{AC1}
\end{equation}
on the metric (\ref{TNEUC}) yields

\begin{equation}
\begin{array}{c}
ds_{(I)}^{2}=\widetilde{V}(r)\left( d\chi +2N\sinh \theta _{1}d\phi
_{1}+2N\cos \theta _{2}d\phi _{2}\right) ^{2}+\frac{dr^{2}}{\widetilde{V}(r)}%
+(r^{2}+N^{2}) \\ 
\times \left( -d\theta _{1}^{2}+\cosh ^{2}\theta _{1}d\phi _{1}^{2}+d\theta
_{2}^{2}-\sin ^{2}\theta _{2}d\phi _{2}^{2}\right)
\end{array}
\label{TN6B1}
\end{equation}
where

\begin{equation}
\widetilde{V}(r)=\frac{1}{3\ell ^{2}(r^{2}+N^{2})^{2}}[3r^{6}+(\ell
^{2}-15N^{2})r^{4}-3N^{2}(2\ell ^{2}-15N^{2})r^{2}-6mr\ell ^{2}-3N^{4}(\ell
^{2}-15N^{2})]  \label{VTNB1}
\end{equation}
To find the structure of the bubble, we set $r=r_{B\text{ }},$ where $%
\widetilde{V}(r_{B})=0.$ Then the induced metric on the bubble is

\begin{equation}
ds_{B(I)}^{2}=(r_{B}^{2}+N^{2})[-d\theta _{1}^{2}+\cosh ^{2}\theta _{1}d\phi
_{1}^{2}+d\theta _{2}^{2}-\sin ^{2}\theta _{2}d\phi _{2}^{2}]  \label{nut1}
\end{equation}
Note that the angle $\theta _{2}$ will be restricted to within a certain
range in order for $\partial /\partial \phi _{2}$ to be a timelike Killing
vector.

Another continuation is

\begin{equation}
(\phi _{1},\phi _{2},n)\rightarrow i(-\phi _{1},-\phi _{2},N)  \label{AC3}
\end{equation}
and this gives

\begin{equation}
\begin{array}{c}
ds_{(III)}^{2}=\widetilde{V}(r)\left( d\chi +2N\cos \theta _{1}d\phi
_{1}+2N\cos \theta _{2}d\phi _{2}\right) ^{2}+\frac{dr^{2}}{\widetilde{V}(r)}%
+(r^{2}+N^{2}) \\ 
\times \left( d\theta _{1}^{2}-\sin ^{2}\theta _{1}d\phi _{1}^{2}+d\theta
_{2}^{2}-\sin ^{2}\theta _{2}d\phi _{2}^{2}\right)
\end{array}
\label{TN6B3}
\end{equation}
with the induced bubble metric:

\begin{equation}
ds_{B(III)}^{2}=(r_{B}^{2}+N^{2})[d\theta _{1}^{2}-\sin ^{2}\theta _{1}d\phi
_{1}^{2}+d\theta _{2}^{2}-\sin ^{2}\theta _{2}d\phi _{2}^{2}]  \label{nut3}
\end{equation}
Similarly, the following continuation:

\begin{equation}
(\theta _{1},\theta _{2},\phi _{1},\phi _{2},n)\rightarrow i(\theta
_{1},\theta _{2},-\phi _{1},-\phi _{2},N)  \label{AC4}
\end{equation}
gives

\begin{equation}
\begin{array}{c}
ds_{(IV)}^{2}=\widetilde{V}(r)\left( d\chi +2N\cosh \theta _{1}d\phi
_{1}+2N\cosh \theta _{2}d\phi _{2}\right) ^{2}+\frac{dr^{2}}{\widetilde{V}(r)%
}+(r^{2}+N^{2}) \\ 
\times \left( -d\theta _{1}^{2}+\sinh ^{2}\theta _{1}d\phi _{1}^{2}-d\theta
_{2}^{2}+\sinh ^{2}\theta _{2}d\phi _{2}^{2}\right)
\end{array}
\label{TN6B4}
\end{equation}
with the induced bubble metric:

\begin{equation}
ds_{B(IV)}^{2}=(r_{B}^{2}+N^{2})[-d\theta _{1}^{2}+\sinh ^{2}\theta
_{1}d\phi _{1}^{2}-d\theta _{2}^{2}+\sinh ^{2}\theta _{2}d\phi _{2}^{2}]
\label{nut4}
\end{equation}
which again has two timelike directions.

Similar considerations apply for all higher-dimensional Taub-Nut-AdS
solutions have $U(1)$ fibrations over a product of $S^{2}$'s. For example
the 8-dimensional solutions has a $U(1)$ fibration over $S^{2}\times
S^{2}\times S^{2}$ \cite{awad}: 
\begin{equation}
\begin{array}{c}
ds^{2}=F(r)\left( d\chi +2n\cos \theta _{1}d\phi _{1}+2n\cos \theta
_{2}d\phi _{2}+2n\cos \theta _{3}d\phi _{3}\right) ^{2}+\frac{dr^{2}}{F(r)}%
+(r^{2}-n^{2}) \\ 
\times \left( d\theta _{1}^{2}+\sin ^{2}\theta _{1}d\phi _{1}^{2}+d\theta
_{2}^{2}+\sin ^{2}\theta _{1}d\phi _{2}^{2}+d\theta _{3}^{2}+\sin ^{2}\theta
_{3}d\phi _{3}^{2}\right)
\end{array}
\label{TNADS8}
\end{equation}
where 
\begin{equation}
F(r)=\frac{(r-n)(5r^{4}+20nr^{3}+(\ell ^{2}+22n^{2})r^{2}+(4n\ell
^{2}-12n^{3})r-35n^{4}+5\ell ^{2}n^{2}}{5(r+n)^{3}\ell ^{2}}  \label{FTN8}
\end{equation}
In this case, the mass parameter $m$ takes the value 
\begin{equation}
m=\frac{8n^{5}(\ell ^{2}-8n^{2})}{5\ell ^{2}}  \label{MTN8}
\end{equation}
to have a nut solution. The coordinate $\tau $ has a period $12\pi n$, to
avoid a conical singularity at $r=n.$ Similarly, the ten dimensional
Taub-Nut-AdS solution is a result of a $U(1)$ fibration over $S^{2}\times
S^{2}\times S^{2}\times S^{2}$ $:$ 
\begin{equation}
\begin{array}{c}
ds^{2}=F(r)\left( d\chi +2n\cos \theta _{1}d\phi _{1}+2n\cos \theta
_{2}d\phi _{2}+2n\cos \theta _{3}d\phi _{3}+2n\cos \theta _{4}d\phi
_{4}\right) ^{2}+\frac{dr^{2}}{F(r)}+(r^{2}-n^{2}) \\ 
\times \left( d\theta _{1}^{2}+\sin ^{2}\theta _{1}d\phi _{1}^{2}+d\theta
_{2}^{2}+\sin ^{2}\theta _{1}d\phi _{2}^{2}+d\theta _{3}^{2}+\sin ^{2}\theta
_{3}d\phi _{3}^{2}+d\theta _{4}^{2}+\sin ^{2}\theta _{4}d\phi _{4}^{2}\right)
\end{array}
\label{TN10}
\end{equation}
where for a NUT solution, the mass parameter and function $F(r)$ are given
by: 
\begin{equation}
m=\frac{64n^{7}(10n^{2}-\ell ^{2})}{35\ell ^{2}}  \label{MTN10}
\end{equation}
and 
\begin{equation}
\begin{array}{c}
F(r)=\frac{r-n}{35(r+n)^{4}\ell ^{2}}\{35r^{5}+175nr^{4}+(300n^{2}+5\ell
^{2})r^{3}+(25n\ell ^{2}+100n^{3})r^{2} \\ 
+(47n^{2}\ell ^{2}-295n^{4})r-315n^{5}+35\ell ^{2}n^{3}\}
\end{array}
\label{FTN10}
\end{equation}

The coordinate $\tau $ must has a period $20\pi n$, to avoid a conical
singularity at $r=n.$

The form for a general fibration over $\left( S^{2}\right) ^{\otimes k}$ is
the same. All analytic continuations in the $S^{2}$ sections force the
introduction of more than one timelike direction because the NUT parameter
must be analytically continued to avoid a complex metric.

Another possibility is a over ${\Bbb CP}^{2}$. In six dimensions this is 
\cite{awad}: 
\begin{equation}
ds^{2}=V(r)\left( d\chi +nA\right) ^{2}+\frac{dr^{2}}{V(r)}%
+(r^{2}-n^{2})d\Sigma _{2}^{2}  \label{TN6CP2}
\end{equation}
where $A$ is given by 
\begin{equation}
A=\frac{u^{2}}{2(1+\frac{u^{2}}{6})}(d\psi +\cos \theta d\phi )  \label{A6}
\end{equation}
and $d\Sigma _{2}^{2}$ is the metric over the ${\Bbb CP}^{2}$, which is
given by 
\begin{equation}
d\Sigma _{2}^{2}=\frac{1}{(1+\frac{u^{2}}{6})^{2}}\{du^{2}+\frac{u^{2}}{4}%
(d\psi +\cos \theta d\phi )^{2}\}+\frac{u^{2}}{4(1+\frac{u^{2}}{6})}(d\theta
^{2}+\sin ^{2}\theta d\phi ^{2})  \label{CP2metric}
\end{equation}

The function $V(r)$ in the metric is the same as function (\ref{VTNEUC}),
appearing in the Taub-Nut-AdS fibered over $S^{2}\times S^{2}.$ In the limit 
$\ell \rightarrow \infty ,$ the metric (\ref{TN6CP2}) approaches the metric
considered in \cite{Bais}. The situation here is somewhat different in that
we have not found any analytic continuation of the coordinates of the ${\Bbb %
CP}^{2}$ section that yields a single timelike direction. \ Further
higher-dimensional generalizations are products of $S^{2}$ and ${\Bbb CP}%
^{2} $ and the same considerations apply, namely more than one timelike
direction appears after analytic continuation unless $\chi $ is the
coordinate continued.

\section{Kerr-Nut Bubbles}

\begin{figure}[tbp]
\begin{center}
{\large \epsfig{file=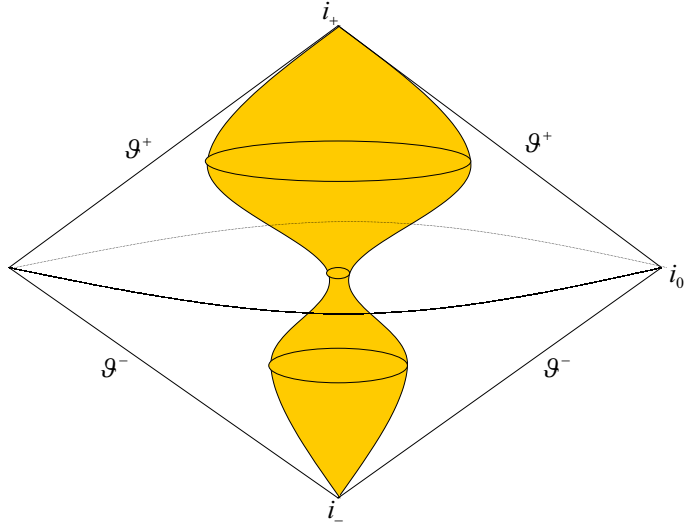,width=0.6\linewidth} }
\end{center}
\caption{The Penrose diagram of the Kerr-Nut bubble (\ref{KN1}) with
coordinates $(t,\tilde{\protect\chi})$. The space inside the shaded region
is absent and the surface of that region is the bubble.}
\label{fig4}
\end{figure}

We consider next rotating NUT solutions in four dimensions. Such solutions
are given by the class of Kerr-Nut metrics, which have the (Euclidean) form 
\cite{Mannnutrot} 
\begin{equation}
\begin{array}{c}
ds^{2}=\frac{V(r)[d\chi -(2n\cos \theta -a\sin ^{2}\theta )d\phi ]^{2}+\sin
^{2}\theta \lbrack ad\chi -(r^{2}-n^{2}-a^{2})d\phi ]^{2}}{[r^{2}-(n+a\cos
\theta )^{2}]} \\ 
+[r^{2}-(n+a\cos \theta )^{2}](\frac{dr^{2}}{V(r)}+d\theta ^{2})
\end{array}
\label{KN}
\end{equation}
where 
\begin{equation}
V(r)=r^{2}-2mr-a^{2}+n^{2}  \label{knutV}
\end{equation}
Applying the following analytic continuations

\begin{equation}
(\theta ,n)\rightarrow (it+\frac{\pi }{2},iN)  \label{AAc1KN}
\end{equation}
on the metric (\ref{KN}) yields: 
\begin{equation}
\begin{array}{c}
ds_{1}^{2}=\frac{\widetilde{V}(r)[d\chi -(2N\sinh t-a\cosh ^{2}t)d\phi
]^{2}+\cosh ^{2}t[ad\chi -(r^{2}+N^{2}-a^{2})d\phi ]^{2}}{[r^{2}+(N-a\sinh
t)^{2}]} \\ 
+[r^{2}+(N-a\sinh t)^{2}](\frac{dr^{2}}{\widetilde{V}(r)}-dt^{2})
\end{array}
\label{KN1}
\end{equation}
where the function $\widetilde{V}(r)$ $\ $is given by: 
\begin{equation}
\widetilde{V}(r)=r^{2}-2mr-a^{2}-N^{2}  \label{VKN1}
\end{equation}
The largest root of $\widetilde{V}(r)$ is $\widetilde{r}_{+}=m+\sqrt{%
m^{2}+a^{2}+N^{2}}$ and so the induced metric on the bubble is 
\begin{equation}
ds_{b}^{2}=-[\widetilde{r}_{+}^{2}+(N-a\sinh t)^{2}]dt^{2}+\frac{\cosh ^{2}td%
\tilde{\chi}^{2}}{[\widetilde{r}_{+}^{2}+(N-a\sinh t)^{2}]}  \label{KN1I}
\end{equation}
where $\tilde{\chi}=a\chi -(\widetilde{r}_{+}^{2}+N^{2}-a^{2})\phi $. For
all values of $t,$ the coefficient of time is negative definite. So, unlike
the case of Kerr-AdS \cite{Bir}, the induced bubble metric obtained through
the analytic continuation of the Kerr-Nut metric in four dimension has a
well defined behaviour. In the special case of $\ N=0,$ (\ref{KN1I}) reduces
to the Kerr bubble obtained in \cite{Aha}. For $(N-a\sinh t)^{2}<<\widetilde{%
r}_{+}^{2},$ the bubble evolves like dS$_{2}$. This situation is similar to
the case of Schwarzschild and Kerr bubbles \cite{Aha}. However the presence
of a NUT charge causes the time evolution to become asymmetric, with the
bubble approaching a minimal radius and then later a maximal radius near $%
t=0 $ (or the time-reversal of this, depending on the signs of the
parameters). For very large $t,$ the bubble metric becomes 
\begin{equation}
ds_{b}^{2}=-\frac{1}{4}a^{2}e^{2t}dt^{2}+\frac{d\tilde{\chi}^{2}}{a^{2}}
\label{KN1Ia}
\end{equation}
which is independent of the NUT charge. In this limit, the circle
parametrized by $\tilde{\chi}$ remains at a fixed radius, so we do not have
any time evolution for the bubble. So when the bubble evolves like dS$_{2},$
the effect of nut charge is to shift the value of the coordinate $t$ to
satisfy the above mentioned inequality. On the other hand, for large values
of $t,$ the NUT charge does not have any physical effect on the induced
bubble metric (\ref{KN1I}), and the spacetime evolves like a Milne universe
at late times. The{\large \ }Penrose diagram of the Kerr-Nut bubble (\ref
{KN1I}) is given in figure (\ref{fig4}). 
For the four dimensional Kerr-Nut bubble (\ref{KN1}), the null geodesics
relevant for the s-waves satisfy: 
\begin{equation}
\frac{dr^{2}}{\widetilde{V}(r)}=dt^{2}  \label{null}
\end{equation}
Integrating the above equation, we find the relation 
\begin{equation}
e^{t}=A(r-m\pm \widetilde{V}(r))=A(r-m\pm \sqrt{r^{2}-2mr-a^{2}-N^{2}})
\label{trnull}
\end{equation}
where $A$\ is a constant labeling the different null geodesics and $\pm $\
refers to outgoing and incoming modes. An s-wave at $\vartheta ^{+}$\ is
given by a purely positive frequency $e^{i\omega U}$, where $U=T-X.$\ Here $%
T $\ and $X$\ are the Minkowski coordinates in the time and radial
directions at infinity. For this s-wave, the phase difference between two
geodesics at\ $U_{1}$\ and $U_{2}$\ on $\vartheta ^{+},$\ i.e. $\Delta U,$\
can be related to the phase difference traced back to $V_{1}$\ and $V_{2}$\
on $\vartheta ^{-},$\ where $V=T+X.$\ The bubble metric in the $(t,r)$\
section for very large $r$ (such that $r\gg a,N,m$) and very large $t,$\ ($%
t\rightarrow \pm \infty $), is given by

\begin{equation}
ds^{2}=-dUdV  \label{KN1UV}
\end{equation}
where in terms of the original coordinates 
\begin{equation}
U=-re^{-t}+\frac{a^{2}e^{t}}{4r},V=re^{t}-\frac{a^{2}e^{-t}}{4r}  \label{UV}
\end{equation}
Using relation (\ref{trnull}), we find that the future coordinate $U$\ of
the null geodesic is 
\begin{equation}
U\simeq \frac{a^{2}}{2}A-\frac{1}{2A}  \label{UU}
\end{equation}
\ Similarly, the past coordinate $V$\ is 
\begin{equation}
V\simeq (\frac{a^{2}+N^{2}}{2}-m^{2})A-\frac{a^{2}}{4A}\frac{1}{\frac{%
a^{2}+N^{2}}{2}-m^{2}}  \label{VV}
\end{equation}
As with the $N=0$ case \cite{Aha}, we can find $\Delta U$\ between two
geodesics. We solve the equation (\ref{VV})\ for $A_{2\text{ }}$in terms of $%
V_{2}$\ (taking geodesic $1$\ as the reference geodesic) and then substitute
the result in eq.(\ref{UU}). This gives 
\begin{equation}
\Delta U=\frac{a^{2}}{2}(A_{1}-\frac{V_{2}+\sqrt{V_{2}^{2}+a^{2}}}{%
a^{2}+N^{2}-2m^{2}})-\frac{1}{2}(\frac{1}{A_{1}}-\frac{a^{2}+N^{2}-2m^{2}}{%
V_{2}+\sqrt{V_{2}^{2}+a^{2}}})  \label{DELTAU}
\end{equation}
indicating that a positive frequency s-wave at future can not continue back
to a purely positive frequency wave at past and so this means that we have a
particle creation in bubble spacetime (\ref{KN1}). From the above obtained
result, one can calculate the Bogoliubov coefficients (which determines the
extent of particle creation), the total number of particles and other
qualitative results about the particle creation.

\begin{figure}[tbp]
\begin{center}
\epsfig{file=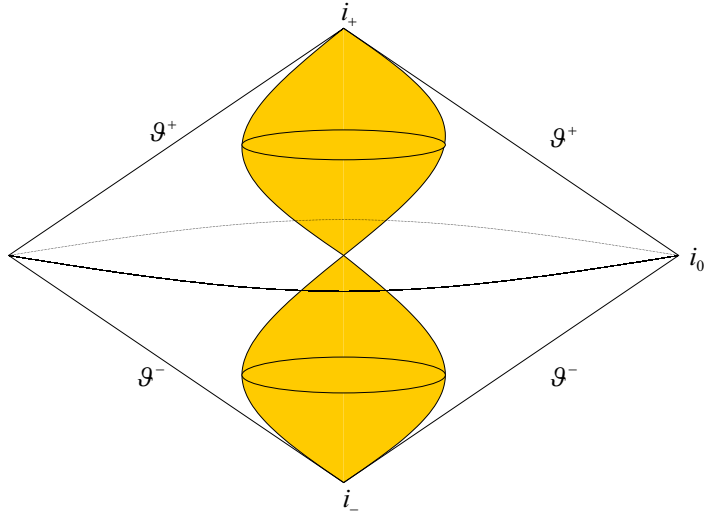,width=0.6\linewidth}
\end{center}
\caption{The Penrose diagram of \ the Kerr-Nut bubble (\ref{KN2}) with
coordinates $(t,\widehat{\protect\chi })$. The space inside the shaded
region is absent and the surface of that region is the bubble.}
\label{fig5}
\end{figure}

Another analytic continuation of the metric (\ref{KN}) is obtained through 
\begin{equation}
(\theta ,\phi ,n,a)\rightarrow i(t,\phi ,N,a)  \label{AAC2KN}
\end{equation}
which yields: 
\begin{equation}
\begin{array}{c}
ds_{2}^{2}=\frac{V^{\ast }(r)[d\chi +(2N\cosh t+a\sinh ^{2}t)d\phi
]^{2}+\sinh ^{2}t[ad\chi -(r^{2}+N^{2}+a^{2})d\phi ]^{2}}{[r^{2}+(N+a\cosh
t)^{2}]} \\ 
+[r^{2}+(N+a\cosh t)^{2}](\frac{dr^{2}}{V^{\ast }(r)}-dt^{2})
\end{array}
\label{KN2}
\end{equation}
where the function $V^{\ast }(r)$ is given by: 
\begin{equation}
V^{\ast }(r)=r^{2}-2mr+a^{2}-N^{2}  \label{VKN2}
\end{equation}
In this case $r=r_{+}^{\ast }$ $=m+\sqrt{m^{2}+N^{2}-a^{2}}$, the induced
metric on the bubble is given by: 
\begin{equation}
ds_{b}^{2}=-[r_{+}^{\ast 2}+(N+a\cosh t)^{2}]dt^{2}+\frac{\sinh ^{2}td%
\widehat{\chi }^{2}}{[r_{+}^{\ast 2}+(N+a\cosh t)^{2}]}  \label{KN2I}
\end{equation}
where $\widehat{\chi }=a\chi -(r_{+}^{\ast 2}+N^{2}+a^{2})\phi .$ So again
we have a bubble solution. For $(N+a\cosh t)^{2}<<r_{+}^{\ast 2},$ the
bubble evolves like dS$_{2}$, now with symmetric time evolution. \ However
in this case there are two possible time-evolution scenarios. \ If $N$ and $%
a $ are both of the same sign, the bubble has only one minimal vanishing
radius. If $N$ and $a$ have opposite sign then the bubble will expand to a
maximum, shrink to zero, expand again to the same maximum before shrinking
once again. For large $t,$ the bubble metric becomes 
\begin{equation}
ds_{b}^{2}=-\frac{1}{4}a^{2}e^{2t}dt^{2}+\frac{d\widehat{\chi }^{2}}{a^{2}}
\label{KN2Ia}
\end{equation}
again, independent of the NUT charge. Hence in this limit, the circle
parametrized by $\widehat{\chi }$ remains at a fixed radius, so we do not
have any time evolution for the bubble. As with the last case, when the
bubble evolves like dS$_{2},$ the effect of the NUT charge is to shift the
value of coordinate $t$ to satisfy the above mentioned inequality. On the
other hand, for large values of coordinate $t,$ the nut charge does not have
any physical effect on the induced bubble metric (\ref{KN2I}), whose
late-time evolution is also Milne-like. A typical Penrose diagram of the
Kerr-Nut bubble (\ref{KN2I}), when $N$\ and $a$\ have the same sign, is
given by figure (\ref{fig5}). Also, by considering an s-wave in the bubble
background (\ref{KN2}), we can show that particle creation can occur. 

\begin{figure}[tbp]
\begin{center}
\epsfig{file=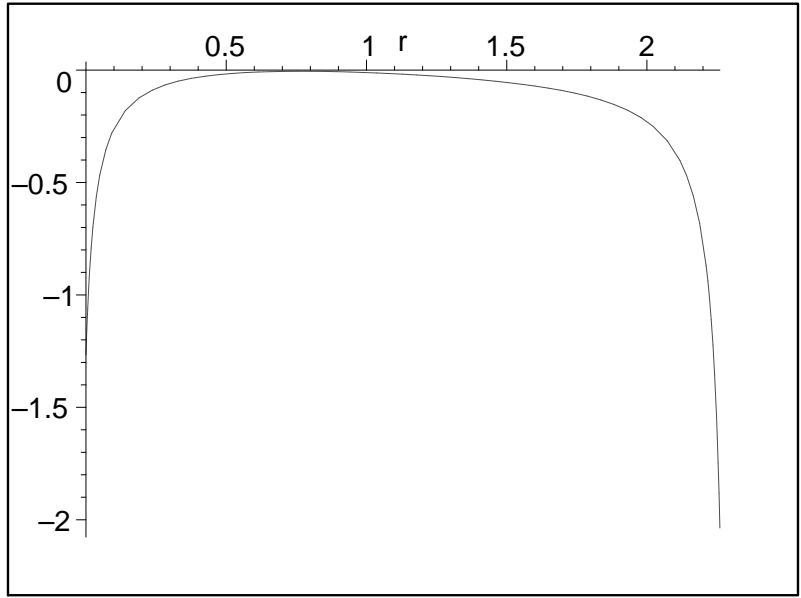,width=0.6\linewidth}
\end{center}
\caption{Effective potential (\ref{pot}) in the special case of $a=N=m=P_{%
\protect\tau }=P_{\protect\phi }=1$ on a constant time slice $\protect\theta %
=1.$The curve does not quite meet the $r$-axis at the top.}
\label{fig6}
\end{figure}

We can describe the causal structure of Kerr-NUT bubbles (\ref{KN1}) and (%
\ref{KN2}) as follows. If we consider two observers at different coordinates 
$r,$\ they are causally connected even for times $t\rightarrow \pm \infty .$%
\ To see this, we consider a null curve which satisfies 
\begin{eqnarray}
\frac{V_{0}(r)[\stackrel{\cdot }{\chi }+(2N\cosh t+a\sinh ^{2}t)\stackrel{%
\cdot }{\phi }]^{2}+\sinh ^{2}t[a\stackrel{\cdot }{\chi }-(r^{2}+N^{2}+a^{2})%
\stackrel{\cdot }{\phi }]^{2}}{[r^{2}+(N+a\cosh t)^{2}]} &&  \nonumber \\
+[r^{2}+(N+a\cosh t)^{2}](\frac{\stackrel{\cdot }{r}^{2}}{V_{0}(r)}-%
\stackrel{\cdot }{t}^{2}) &=&0
\end{eqnarray}
where $V_{0}(r)=\widetilde{V}(r)$ or $V^{\ast }(r)$ and the overdot
represents derivative with respect to an affine parameter labeling the
curve. Since we assumed $V_{0}(r)>0,$\ we conclude that 
\begin{equation}
\frac{\stackrel{\cdot }{r}^{2}}{V_{0}(r)}\leq \,\stackrel{\cdot }{t}^{2}
\label{re}
\end{equation}
Hence, for points on the null curve, when time coordinate $t\rightarrow \pm
\infty $, we do not have any restriction on the maximum change in the
coordinate $r$. So, two observers at two points with different values of $r$%
\ can communicate with each other. The two Killing vectors $(\partial
/\partial \phi )$ and $(\partial /\partial \chi )$ yield two constants of
the motion 
\begin{equation}
\begin{array}{c}
P_{\chi }=\frac{V_{0}(r)+a^{2}\sinh ^{2}t}{r^{2}+(N+a\cosh t)^{2}}\stackrel{%
\cdot }{\chi }+\frac{V_{0}(r)(2N\cosh t+a\sinh ^{2}t)-a\sinh
^{2}t(r^{2}+N^{2}+a^{2})}{r^{2}+(N+a\cosh t)^{2}}\stackrel{\cdot }{\phi } \\ 
P_{\phi }=\frac{V_{0}(r)(2N\cosh t+a\sinh ^{2}t)-a\sinh
^{2}t(r^{2}+N^{2}+a^{2})}{r^{2}+(N+a\cosh t)^{2}}\stackrel{\cdot }{\chi }+%
\frac{V_{0}(r)(2N\cosh t+a\sinh ^{2}t)^{2}+\sinh ^{2}t(r^{2}+N^{2}+a^{2})^{2}%
}{r^{2}+(N+a\cosh t)^{2}}\stackrel{\cdot }{\phi }
\end{array}
\label{com}
\end{equation}
We can use the equation (\ref{com}) to eliminate $\stackrel{\cdot }{\chi }$
and $\stackrel{\cdot }{\phi }$ from the null geodesics. In this case, for a
fixed time $t,$ the null geodesic equation reduces to that of motion of a
particle in an effective potential:

\begin{equation}
\frac{1}{2}\stackrel{\cdot }{r}^{2}+{\cal V}(P_{\chi },P_{\phi },r)=0
\label{eqm}
\end{equation}
where the effective potential is: 
\begin{equation}
{\cal V}(P_{\chi },P_{\phi },r)=\frac{g_{\chi \chi }(g_{\phi \phi }P_{\chi
}-g_{\chi \phi }P_{\phi })^{2}+g_{\phi \phi }(g_{\chi \phi }P_{\chi
}-g_{\chi \chi }P_{\phi })^{2}+2g_{\chi \phi }(g_{\phi \phi }P_{\chi
}-g_{\chi \phi }P_{\phi })(g_{\chi \chi }P_{\phi }-g_{\chi \phi }P_{\chi })}{%
2(r^{2}+(N+a\cosh t)^{2})(g_{\chi \chi }g_{\phi \phi }-g_{\chi \phi
}^{2})^{2}}  \label{pot}
\end{equation}
So geodesics exist in the region for which ${\cal V}(P_{\chi },P_{\phi
},r)\leq 0.$ For example, on a constant time slice $t=1,$ if we set $%
a=N=m=P_{\chi }=P_{\phi }=1,$ the effective potential is plotted in figure (%
\ref{fig6}). 
It is infinite at $r_{0}\approx 2.29777.$ For $r>r_{0},$ the potential is
positive definite and so a geodesic cannot penetrate into this region.

\section{Kerr-Nut-AdS Bubbles}

The four dimensional Euclidean Kerr-Nut-AdS spacetime is given by \cite
{Mannnutrot} : 
\begin{equation}
\begin{array}{c}
ds^{2}=\frac{V(r)[d\chi -(2n\cos \theta -a\sin ^{2}\theta )d\phi
]^{2}+H(\theta )\sin ^{2}\theta \lbrack ad\chi -(r^{2}-n^{2}-a^{2})d\phi
]^{2}}{\sigma ^{4}[r^{2}-(n+a\cos \theta )^{2}]} \\ 
+[r^{2}-(n+a\cos \theta )^{2}](\frac{dr^{2}}{V(r)}+\frac{d\theta ^{2}}{%
H(\theta )})
\end{array}
\label{KERRNUT}
\end{equation}
where the functions $V(r)$ and $H(\theta )$ are given by the following
expressions: 
\begin{equation}
V(r)=\frac{r^{4}}{\ell ^{2}}+\frac{(\ell ^{2}-a^{2}-6n^{2})r^{2}}{\ell ^{2}}%
-2mr-\frac{(a^{2}-n^{2})(\ell ^{2}-3n^{2})}{\ell ^{2}}  \label{VKERRNUT}
\end{equation}
and 
\begin{equation}
H(\theta )=1-4\frac{n^{2}}{\ell ^{2}}+\frac{(2n+a\cos \theta )^{2}}{\ell ^{2}%
}  \label{HKERRNUT}
\end{equation}
There is a bolt solution $r_{b}$ given by $V(r_{b})=0$, with $r_{b}>N$; no\
analogue of the pure NUT solution exists for this case \cite{Mannnutrot}. To
avoid conical singularities in different sections of the spacetime, we must
impose some constraints. For example, avoiding conical singularities in the $%
(\theta ,\phi )$ section implies 
\begin{equation}
\sigma =(1+\frac{a^{2}}{\ell ^{2}})^{-\frac{1}{2}}  \label{chi}
\end{equation}
The period of the coordinate $\chi $ is $\frac{2\pi }{\kappa }=8\pi n,$
where the left and right hand sides of the equality come from avoiding
conical singularities in the $(\chi ,r)$ section and ensuring regularity of
the metric along the $z$-axis. In the previous equality, $\kappa =\frac{%
V^{\prime }(r_{+})}{4\pi \sigma ^{2}(r_{+}^{2}-r_{0}^{2})}$ and $r_{+}=r_{b}$
is the real root of $V(r).$ Applying the following analytic continuations

\begin{equation}
(\theta ,n)\rightarrow (it+\frac{\pi }{2},iN)  \label{AAc1}
\end{equation}
on the metric (\ref{KERRNUT}) yields: 
\begin{equation}
\begin{array}{c}
ds_{1}^{2}=\frac{\widetilde{V}(r)[d\chi -(2N\sinh t-a\cosh ^{2}t)d\phi ]^{2}+%
\widetilde{H}(t)\cosh ^{2}t[ad\chi -(r^{2}+N^{2}-a^{2})d\phi ]^{2}}{\sigma
^{4}[r^{2}+(N-a\sinh t)^{2}]} \\ 
+[r^{2}+(N-a\sinh t)^{2}](\frac{dr^{2}}{\widetilde{V}(r)}-\frac{dt^{2}}{%
\widetilde{H}(t)})
\end{array}
\label{KERRNUT1}
\end{equation}
where the functions $\widetilde{V}(r)$ and $\widetilde{H}(t)$ are given by: 
\begin{equation}
\widetilde{V}(r)=\frac{r^{4}}{\ell ^{2}}+\frac{(\ell ^{2}-a^{2}+6N^{2})r^{2}%
}{\ell ^{2}}-2mr-\frac{(a^{2}+N^{2})(\ell ^{2}+3N^{2})}{\ell ^{2}}
\label{VKERRNUT1}
\end{equation}
and 
\begin{equation}
\widetilde{H}(t)=1+4\frac{N^{2}}{\ell ^{2}}-\frac{(2N-a\sinh t)^{2}}{\ell
^{2}}  \label{HKERRNUT1}
\end{equation}
At $r=\widetilde{r}_{+}$ (where $\widetilde{V}(\widetilde{r}_{+})=0$), and
for the coordinate $t$ in the range of $0\leq $ $t<t_{c},$ the function $%
\widetilde{H}(t)$ is positive definite, and the induced metric on the bubble
is given by: 
\begin{equation}
ds_{b}^{2}=-[\widetilde{r}_{+}^{2}+(N-a\sinh t)^{2}]\frac{dt^{2}}{\widetilde{%
H}(t)}+\frac{\widetilde{H}(t)\cosh ^{2}t[ad\chi -(\widetilde{r}%
_{+}^{2}+N^{2}-a^{2})d\phi ]^{2}}{\sigma ^{4}[\widetilde{r}%
_{+}^{2}+(N-a\sinh t)^{2}]}  \label{KERRNUT1I}
\end{equation}
where the critical coordinate $t_{c}$ is: 
\begin{equation}
t_{c}=\sinh ^{-1}\{\frac{2N}{a}+\frac{\ell }{a}\sqrt{1+\frac{4N^{2}}{\ell
^{2}}}\}  \label{thetacriticalKERRNUT}
\end{equation}
On the other hand, for the range $t>t_{c},$ the function $\widetilde{H}(t)$
is negative definite and so the induced metric (\ref{KERRNUT1I}) does not
describe a well defined bubble. In fact, the situation is similar to the
induced bubble metrics obtained through analytic continuation of the
Kerr-AdS metric in four dimensions \cite{Bir}. In that case the induced
metric also changes signature at some critical time. \ The result of \cite
{Bir} is a special case of our result when $N=0$.

Another special case of the metric (\ref{KERRNUT}) is when $a=0.$ In this
case, the functions $H(t)=\widetilde{H(}t)=\sigma =1$ and from relation (\ref
{thetacriticalKERRNUT}), we have $t_{c}\rightarrow \infty ,$ so we have a
well behaved induced bubble metric 
\begin{equation}
ds_{b}^{2}=(r_{+}^{2}+N^{2})[-dt^{2}+\cosh ^{2}td\phi ^{2}]  \label{NUTI}
\end{equation}
where we derived this form of the induced metric in section one.

The other analytic continuation of the metric (\ref{KERRNUT}) is obtained
through 
\begin{equation}
(\theta ,\phi ,n,a)\rightarrow i(t,\phi ,N,a)  \label{AAC2}
\end{equation}
which yields 
\begin{equation}
\begin{array}{c}
ds_{2}^{2}=\frac{V^{\ast }(r)[d\chi +(2N\cosh t+a\sinh ^{2}t)d\phi
]^{2}+H^{\ast }(t)\sinh ^{2}t[ad\chi -(r^{2}+N^{2}+a^{2})d\phi ]^{2}}{\sigma
^{\ast 4}[r^{2}+(N+a\cosh t)^{2}]} \\ 
+[r^{2}+(N+a\cosh t)^{2}](\frac{dr^{2}}{V^{\ast }(r)}-\frac{dt^{2}}{H^{\ast
}(t)})
\end{array}
\label{KERRNUT2}
\end{equation}
where the functions $V^{\ast }(r)$ and $H^{\ast }(t)$ are given by 
\begin{equation}
V^{\ast }(r)=\frac{r^{4}}{\ell ^{2}}+\frac{(\ell ^{2}+a^{2}+6N^{2})r^{2}}{%
\ell ^{2}}-2mr+\frac{(a^{2}-N^{2})(\ell ^{2}+3N^{2})}{\ell ^{2}}
\label{VKERRNUT2}
\end{equation}
and 
\begin{equation}
H^{\ast }(t)=1+4\frac{N^{2}}{\ell ^{2}}-\frac{(2N+a\cosh t)^{2}}{\ell ^{2}}
\label{HKERRNUT2}
\end{equation}
and $\sigma ^{\ast }=(1-\frac{a^{2}}{\ell ^{2}})^{-\frac{1}{2}}.$ In this
case at point $r=r_{+}^{\ast }$ (for some values of parameters $N,a,\ell ,m$%
, we have $V^{\ast }(r_{+}^{\ast })=0$), and for the coordinate $t$ in the
range of $0\leq $ $t<t_{c}^{\ast },$ the function $H^{\ast }(t)$ is positive
definite, and the induced metric on the bubble is given by: 
\begin{equation}
ds_{b}^{2}=-[r_{+}^{\ast 2}+(N+a\cosh t)^{2}]\frac{dt^{2}}{H^{\ast }(t)}+%
\frac{H^{\ast }(t)\sinh ^{2}t[ad\chi -(r_{+}^{\ast 2}+N^{2}+a^{2})d\phi ]^{2}%
}{\sigma ^{\ast 4}[r_{+}^{\ast 2}+(N+a\cosh t)^{2}]}  \label{KERRNUT2I}
\end{equation}
where the critical coordinate $t_{c}^{\ast }$ is now

\begin{equation}
t_{c}^{\ast }=\cosh ^{-1}\{-\frac{2N}{a}+\frac{\ell }{a}\sqrt{1+\frac{4N^{2}%
}{\ell ^{2}}}\}  \label{thetacriticalKERRNUT2}
\end{equation}
On the other hand, for the range $t>t_{c}^{\ast },$ the function $H^{\ast
}(t)$ is negative definite and so the induced metric (\ref{KERRNUT2I}) does
not describe a well defined bubble.

\section{Conclusion}

Our consideration of bubble spacetimes with NUT charge has yielded several
new candidate four-dimensional time-dependent backgrounds for string theory
whose properties merit further exploration. \ We have found that the
presence of a NUT charge can considerably modify the evolution of the
spacetime, particularly in the topological and rotating cases. The NUT
charge induces the presence of an ergoregion in the latter case, although
for the spacetimes in which this occurs the bubble is not compact. The
bubble (\ref{nutrind}) is particularly intriguing, since its evolution is
Milne-like throughout. In higher dimensions the presence of a NUT charge
appears to forbid the existence of any bubble spacetime, since the double
analytic continuation yields more than one timelike direction. \ 

We close with a few comments about de Sitter spacetime. All of the above
results in the AdS case can, by the simple continuation $\ell \rightarrow
i\ell ,$ be changed to Taub-Nut-dS bubbles. For example the TN-de Sitter
space can be obtained from (\ref{TNLO}), and is given by:

\begin{equation}
ds^{2}=+F(\tau )\left( d\tau +2N\cos \theta d\phi \right) ^{2}-\frac{d\tau
^{2}}{F(\tau )}+(\tau ^{2}+N^{2})\left( d\theta ^{2}+\sin ^{2}\theta d\phi
^{2}\right)  \label{dSNUT}
\end{equation}
where 
\begin{equation}
F(\tau )=\frac{\tau ^{4}-\left( \ell ^{2}-6N^{2}\right) \tau ^{2}-2m\ell
^{2}\tau +N^{2}\left( \ell ^{2}-3N^{2}\right) }{(\tau ^{2}+N^{2})\ell ^{2}}
\label{FTNDSLO}
\end{equation}
These spacetimes are expected to be better behaved, at least in the rotating
case, since the functions $\tilde{H}(t)$ and $H^{\ast }(t)$ in eqs. (\ref
{KERRNUT1}) and (\ref{HKERRNUT2}) respectively need never vanish if the
parameters are appropriately chosen. We leave a study of this case for
future work.

\bigskip

{\Large Acknowledgments}

This work was supported by the Natural Sciences and Engineering Research
Council of Canada.\

\end{document}